\documentclass[floatfix,prd,twocolumn,superscriptaddress,showpacs,amssymb,amsmath,amsfonts,aps,altaffilletter,nofootinbib,nopreprintnumbers,showpacs,longbibliography]{revtex4-1}
\usepackage[latin9]{inputenc}
\usepackage{float}
\usepackage{graphicx}
\usepackage{esint}
\usepackage[colorlinks]{hyperref}

\makeatletter

\@ifundefined{textcolor}{}
{%
 \definecolor{BLACK}{gray}{0}
 \definecolor{WHITE}{gray}{1}
 \definecolor{RED}{rgb}{1,0,0}
 \definecolor{GREEN}{rgb}{0,1,0}
 \definecolor{BLUE}{rgb}{0,0,1}
 \definecolor{CYAN}{cmyk}{1,0,0,0}
 \definecolor{MAGENTA}{cmyk}{0,1,0,0}
 \definecolor{YELLOW}{cmyk}{0,0,1,0}
}

\makeatother

\begin{document}

\title{A stochastic template bank for gravitational wave searches for precessing\\ neutron-star--black-hole
coalescence events}

\author{Nathaniel Indik}
\affiliation{Max-Planck-Institut f\"ur Gravitationsphysik (Albert-Einstein-Institut), Callinstr. 38, D-30167 Hannover, Germany}
\affiliation{Leibniz Universit\"at Hannover, Welfengarten 1-A, D-30167 Hannover, Germany}

\author{K Haris}
\affiliation{Indian Institute of Science Education and Research Thiruvananthapuram, CET Campus, Trivandrum 695016, India}

\author{Tito \surname{Dal Canton}}
\affiliation{Max-Planck-Institut f\"ur Gravitationsphysik (Albert-Einstein-Institut), Callinstr. 38, D-30167 Hannover, Germany}
\affiliation{NASA Postdoctoral Program Fellow, Goddard Space Flight Center, Greenbelt, MD 20771, USA}
\affiliation{Leibniz Universit\"at Hannover, Welfengarten 1-A, D-30167 Hannover, Germany}

\author{Henning Fehrmann}
\affiliation{Max-Planck-Institut f\"ur Gravitationsphysik (Albert-Einstein-Institut), Callinstr. 38, D-30167 Hannover, Germany}
\affiliation{Leibniz Universit\"at Hannover, Welfengarten 1-A, D-30167 Hannover, Germany}

\author{Badri Krishnan}
\affiliation{Max-Planck-Institut f\"ur Gravitationsphysik (Albert-Einstein-Institut), Callinstr. 38, D-30167 Hannover, Germany}
\affiliation{Leibniz Universit\"at Hannover, Welfengarten 1-A, D-30167 Hannover, Germany}

\author{Andrew Lundgren}
\affiliation{Max-Planck-Institut f\"ur Gravitationsphysik (Albert-Einstein-Institut), Callinstr. 38, D-30167 Hannover, Germany}
\affiliation{Leibniz Universit\"at Hannover, Welfengarten 1-A, D-30167 Hannover, Germany}

\author{Alex B Nielsen}
\affiliation{Max-Planck-Institut f\"ur Gravitationsphysik (Albert-Einstein-Institut), Callinstr. 38, D-30167 Hannover, Germany}
\affiliation{Leibniz Universit\"at Hannover, Welfengarten 1-A, D-30167 Hannover, Germany}

\author{Archana Pai}
\affiliation{Indian Institute of Science Education and Research Thiruvananthapuram, CET Campus, Trivandrum 695016, India}

\date{\today}
\begin{abstract}

  Gravitational wave searches to date have largely focused on
  non-precessing systems.  Including precession effects greatly
  increases the number of templates to be searched over. This leads to
  a corresponding increase in the computational cost and can increase
  the false alarm rate of a realistic search.  On the other hand,
  there might be astrophysical systems that are entirely missed by
  non-precessing searches. In this paper we consider the problem of
  constructing a template bank using stochastic methods for 
  neutron-star--black-hole binaries allowing for precession, but with the
  restrictions that the orientation of the total angular momentum of the binary is
  pointing towards the detector and that the neutron-star spin is
  negligible relative to that of the black-hole. We quantify the
  number of templates required for the search, and we explicitly
  construct the template bank.  We show that despite the large number
  of templates, stochastic methods can be adapted to solve the
  problem.  We quantify the parameter space region over which the
  non-precessing search might miss signals.

\end{abstract}
\maketitle

\section{Introduction}

\label{sec:intro}

Binary systems consisting of neutron-stars and black-holes are key targets for
the present generation of gravitational wave detectors such as Advanced LIGO
\cite{aligo} and Advanced Virgo \cite{avirgo}.  The LIGO detectors have to date
observed two such events with high significance, labeled GW150914
\cite{Abbott:2016blz} and GW151226 \cite{Abbott:2016nmj} and a third LVT151012
with lower significance \cite{TheLIGOScientific:2016pea}. All three of these are
binary black-hole coalescence events.  The searches for these events \emph{a
priori} cover a wide range of masses and spins magnitudes, but use only
waveforms for which the spins of the individual compact objects are assumed to
be completely aligned or completely anti-aligned with the orbital angular
momentum \cite{TheLIGOScientific:2016qqj}. Misalignments between the spin and
orbital angular momentum generally causes precession of the orbital plane and
additional modulations of the gravitational waveforms \cite{Apostolatos:1994mx}.
While follow-up studies for accurate parameter estimation do include precession
\cite{PhysRevX.6.041014}, including these waveforms directly in the initial
search pipelines is challenging; the dimensionality of the parameter space to be
searched is increased, implying a significant increase in the total number of
templates. This is challenging not only for computational reasons, but also
because a larger number of independent templates leads to a larger probability
for false alarms.  Nevertheless, if a significant number of NSBH systems in our universe display precessional modulations that
cannot be accurately recovered by spin-aligned templates, the search pipeline could potentially detect more
events if precession effects were to be included \cite{Canton:2014uja}.

For searches based on matched filtering with modeled waveforms, the traditional
method of constructing a template bank was to use the parameter space metric
\cite{Owen:1998dk,Dhurandhar:1992mw} for determining the spacing between
adjacent templates.  This method has been successfully used to search for non-spinning
systems \cite{Cokelaer:2007kx} and has also been applied to aligned-spin
systems \cite{Brown:2012qf,Harry:2013tca}.  For precessing waveforms
however, the parameter space metric is not yet sufficiently well understood for
it to be directly used to place templates.  The main issue is that to place a
lattice of templates, one needs a coordinate system on the parameter space where
the metric is explicitly flat. It is not clear whether such a coordinate system
exists (even in any approximate sense) for the space of precessing waveforms.
In situations where such geometric template placement methods are not available,
stochastic methods are commonly employed \cite{Harry:2009ea,Babak:2008rb}. This
also includes the most recent searches over the first Advanced LIGO observing
run \cite{TheLIGOScientific:2016qqj,TheLIGOScientific:2016pea}.

The basic idea of stochastic
methods is to place templates at random points in the parameter space
and to remove templates which happen to lie very close to other
templates.  These stochastic methods are generally applicable but they
are typically less efficient than the geometric methods, i.e. they
require more templates than a geometric bank to achieve the same
coverage over the same parameter space (however, stochastic methods
become more efficient in higher dimensions and can be competitive with
geometric methods \cite{Messenger:2008ta}).  Moreover, the stochastic
template placement procedure can be computationally demanding for
large parameter spaces, which is in fact the case for precessing
waveforms.

In this paper we shall meet this computational challenge
and show how stochastic methods can be applied to cover the space of
precessing waveforms.  The main computational problem we face is that
for every proposed template, one typically compares it with \emph{all}
previously accepted templates to decide whether or not it should be
accepted.  We shall see that with an appropriate choice of
coordinates, it is possible to break up the parameter space into
smaller regions, and treat each region independently.  This paper
presents the largest template bank constructed thus far with
stochastic methods for binary coalescence searches.

Specifically, we shall focus
on neutron-star--black-hole (NSBH) systems, but we expect that our
method would apply to other source systems as well.  We shall
consider NSBH binaries with a black-hole of mass $M_1$ and a neutron
star mass of $M_2$ such that $2M_\odot < M_1 < 16M_\odot$, and
$1M_\odot < M_2 < 3M_\odot$.  Since neutron-star spins are expected to
be small we shall ignore them, but the black-hole spin will be allowed
to take any magnitude which is meaningful in the Kerr metric and any
direction \cite{Lorimer:2008se}.
We shall use the frequency domain waveform model presented in
\cite{Lundgren:2013jla}.  This waveform model does not, so far,
contain the merger and ringdown portions.  For the parameter space
above, and for the expected sensitive frequency range of the Advanced
LIGO and Virgo detectors, the inspiral portion of the waveform will
have the largest contribution to the signal-to-noise ratio.  Thus, the
merger and ringdown phases will not be important for our purposes.
However, the methods used in our study should be useful also for
higher mass systems where merger effects are more important.

The most effectual implementation of a stochastic NSBH template bank
constructed to date \cite{Harry:2016ijz} required approximately 1.6
million, assuming the detector to be in the ``early Advanced LIGO''
configuration \cite{Aasi:2013wya}.  This construction used a new detection statistic based on maximizing the
signal-to-noise ratio (SNR) over source locations in the sky
and required a minimal match criteria of 90\%
when comparing each proposed template with previously accepted ones,
as opposed to the more conventional 97\%.  Using the conventional 97\%
value would lead to a much larger number of templates.  Moreover, as the 
detector improves its low frequency sensitivity over the next few years,
the number of templates increases further. The method used in this paper 
could be used to deal with both of the above issues.  We shall use the conventional 97\% 
minimal match value and, for simplicity, we use the conventional SNR
rather than the detection statistic introduced in \cite{Harry:2016ijz},
but we expect that our method can be adapted to that detection
statistic as well. 

The plan for the rest of the paper is as follows.
Sec. \ref{sec:background} briefly sets up notation and the parameters
describing a precessing binary system and the gravitational waveform,
and outlines the stochastic template placement algorithm.
Sec. \ref{sec:bank} describes the stochastic template bank.
Sec.~\ref{sec:effectualness} compares this precessing template bank with the aligned spin bank and studies how well it recovers injected signals.

\section{Background}
\label{sec:background}

\subsection{Precessing binaries}
\label{sec:PrecBinaries}

Consider an NSBH system consisting of a black-hole with mass $M_1$,
spin $\mathbf{S}$, and a neutron star of mass $M_2$ and zero spin.  Let
$\mathbf{\hat{N}}(\theta,\phi)$ be the unit vector along the line-of-sight from the
detector to the binary system, and let $\mathbf{L}$ be the orbital
angular momentum of the binary.  Define the dimensionless spin of the black-hole as
$\chi = |\mathbf{S}|/M_1^2$.  The component of $\mathbf{S}$ along
$\mathbf{L}$ will be determined by the quantity
$\kappa = \mathbf{\hat{S}}\cdot\mathbf{\hat{L}}$, and the component of
$\mathbf{S}$ orthogonal to $\mathbf{L}$ is
\begin{equation}
  \mathbf{S^{\mathbf{\bot}}}=\mathbf{S}-(\mathbf{S}\cdot\mathbf{\hat{L}})\mathbf{\hat{L}}\label{eq:S1P}  \,.
\end{equation}
It can be shown that the direction of the total angular momentum
$\mathbf{J}= \mathbf{L}+\mathbf{S}$ is approximately conserved
\cite{Apostolatos:1994mx}, and that $\mathbf{\hat{L}}$ and
$\mathbf{\hat{S}}$ precess around $\mathbf{J}$.  The magnitude of
$\mathbf{L}$ decreases steadily because of the emission of
gravitational radiation but the magnitude of $\mathbf{S}$ remains
constant as does the angle between $\mathbf{L}$ and $\mathbf{S}$.  The
opening angle $\beta\,\,$ of the precession cone is given by
\begin{equation}
  \cos{\beta}\,\,=\mathbf{\hat{J}}\cdot\mathbf{\mathbf{\hat{L}}}\label{eq:beta}  \,.  
\end{equation}
As the magnitude of $\mathbf{L}$ decreases, $\beta\,$ should increase in
order to maintain the direction of $\mathbf{J}$ and the angle between
$\mathbf{L}$ and $\mathbf{S}$ \cite{Apostolatos:1994mx}.  However, the
precession time-scale is smaller than the radiation reaction time scale
(which determines the rate at which $L$ decreases).  It can be shown
\cite{Brown:2012gs} that for the advanced LIGO and Virgo detectors, to
a reasonable approximation, $\mathbf{L}$ and $\mathbf{S}$ precess
steadily around $\mathbf{J}$ with a constant opening angle $\beta\,$.
The rare case of transitional precession occurs when $\mathbf{J} \sim 0$
at some point during the evolution of the binary system.  Finally,
$\alpha_0$ is an azimuthal angle  that expresses the orientation of 
$\mathbf{\hat{L}}$ relative to $\mathbf{\hat{J}}$ in the inertial detector frame and we shall define the angle
$\theta_J$ as $\cos\theta_J = \mathbf{\hat{J}}\cdot \mathbf{\hat{N}}$.

For a plane gravitational wave traveling in a direction
$\mathbf{\hat{z}}$, and a frame $(\mathbf{\hat{x}},\mathbf{\hat{y}})$
in the plane orthogonal to $\mathbf{\hat{z}}$, we define the tensors
\begin{equation}
  \mathbf{e}^+_{ab} = \mathbf{\hat{x}}_a\mathbf{\hat{x}}_b - \mathbf{\hat{y}}_a\mathbf{\hat{y}}_ b  
  \,,\quad \mathbf{e}^\times_{ab} = \mathbf{\hat{x}}_a\mathbf{\hat{y}}_b + \mathbf{\hat{y}}_a\mathbf{\hat{x}}_ b\,.
\end{equation}
The gravitational wave can be written as a sum of two transverse polarizations
\begin{equation}
  h_{ab}(t) = h_+(t)\mathbf{e}^+_{ab} + h_\times(t)\mathbf{e}^\times_{ab}\,.
\end{equation}
It is always possible to find a frame
$(\mathbf{\hat{x}}, \mathbf{\hat{y}})$ such that 
\begin{equation}
  h_+(t) = A_+(t)\cos 2\Phi(t) \,,\qquad h_\times(t) =A_\times(t)\sin 2\Phi(t)\,,
\end{equation}
where $A_{+,\times}$ are slowly varying amplitudes and $\Phi(t)$ is a
rapidly varying phase.  For the case of a binary system, the
wave-frame $(\mathbf{\hat{x}},\mathbf{\hat{y}})$ is tied to the
direction of the orbital angular momentum, and $\mathbf{\hat{x}}$ is
taken to be $\pm \mathbf{\hat{N}}\times\mathbf{\hat{L}}$.  The
direction of $\mathbf{x}$ in the detector frame defines a polarization
angle $\psi$ and, following \cite{Apostolatos:1994mx}, we choose the
convention: 
\begin{equation}
  \psi(t)=\tan^{-1}\left(\frac{\mathbf{\hat{L}}(t)\cdot\mathbf{\hat{z}}-(\mathbf{\hat{L}}(t)\cdot\mathbf{\hat{N}})(\mathbf{\hat{z}}\cdot\mathbf{\hat{N}})}{\mathbf{\hat{N}}\cdot(\mathbf{\hat{L}}(t)\times\mathbf{\hat{z}})}\right)\,.\label{eq:GWphase3}
\end{equation}
Note that because of precession, the direction of $\mathbf{L}$ changes
in time and thus $\psi$ also changes with time.  With these
conventions, the expressions for $h_{+,\times}$ are:
\begin{eqnarray}
h_{+}(t)&=&-\frac{2\pi M}{rD}\left[1+(\mathbf{\hat{L}}(t)\cdot\mathbf{\hat{N}})^{2}\right]\cos2\Phi(t)\label{eq:h_plus}\,,\\
h_{\times}(t)&=&-\frac{2\pi M}{rD}\left[-2~\mathbf{\hat{L}}(t)\cdot\mathbf{\hat{N}}\right]\sin2\Phi(t)\label{eq:h_cross}\,,
\end{eqnarray}
where $D$ is the distance to the binary system, $r$ is the binary orbital diameter
and $M=M_1+M_2$ is the total mass. Throughout the paper we shall also use the
\emph{chirp mass},
\begin{equation}
  \mathcal{M_{\mathrm{C}}} = \eta^{\frac{3}{5}}M\label{eq:Mchirp}\,,
\end{equation}
and the following quantities related to the mass ratio of the components,
\begin{equation}
  \nu = \frac{M_{1}}{M}\, \quad \eta = \frac{M_1 M_2}{M^2}\, \quad q = \frac{M_1}{M_2}\,.
  \label{eq:mu}
\end{equation}

It is also convenient to express the black-hole spin via the dimensionless vector
$\mathbf{\chi} := \mathbf{S}/M_1^2$ and decompose it into components parallel
and perpendicular to $\widehat{\mathbf{L}}$, $\chi^\parallel$ and $\chi^\perp$
respectively. The total dimensionless spin magnitude is thus
$\chi = \sqrt{{(\chi^\parallel)}^2 + {(\chi^\perp)}^2}$.

The detector response functions to these polarizations are denoted by
$F_{+}(\mathbf{N},\psi)$ and $F_{\times}(\mathbf{N},\psi)$. If the
signal is parallel to one of these configurations, it is said to be
linearly polarized. In contrast, a signal that can be decomposed into
an equal linear combination of these two principal directions is
circularly polarized. In general, the signal seen by the detector
$h(t)$ will be a linear combination of the two polarizations:
\begin{eqnarray}
h(t)  &=& h_{+}(t)F_{+}(\mathbf{\hat{N}},\psi(t))+h_{\times}(t)F_{\times}(\mathbf{\hat{N}},\psi(t)) \nonumber \\
      &=& A(t)\cos[2\Phi(t)+\varphi(t)] \label{eq:h_alltogether}\,,
\end{eqnarray}
where
\begin{eqnarray}
A(t) & = & \frac{2\pi M}{rD}\left(\left[1+(\mathbf{\hat{L}}(t)\cdot\mathbf{\hat{N}})^{2}\right]^{2}F_{+}^{2}(\theta,\phi,\psi(t))\right.\nonumber \\
     &  & \left.+ 4~[\mathbf{\hat{L}}(t)\cdot\mathbf{\hat{N}}]^{2}F_{\times}^{2}(\theta,\phi,\psi(t))\right)^{1/2}\,,\label{GWamplitude}
\end{eqnarray}
and
\begin{equation}
\varphi(t)= \tan^{-1}\left(\frac{2~ (\mathbf{\hat{L}}(t)\cdot\mathbf{\hat{N}})~ F_{\times}(\theta,\phi,\psi(t))}{[1+(\mathbf{\hat{L}}(t)\cdot\mathbf{\hat{N}})]^{2}~F_{+}(\theta,\phi,\psi(t))}\right)\,.\label{eq:GWphase2}
\end{equation}
In summary gravitational wave signals from an NSBH precessing binary
system can be expressed in terms of the following parameters: the
component masses $(M_1,M_2)$, the black-hole spin vector $\mathbf{S}$,
the overall constant amplitude $A$, the polar angles of total angular
momentum vector $(\theta_J,\psi_J)$, the location of the source
$(\theta,\phi)$, the time of arrival of the signal $t_0$ and the
initial phase $\phi_0$.  For the purposes of this paper we have chosen
to focus on ``face on'' systems, i.e.  we assume that $\mathbf{J}$ is
either aligned or anti-aligned with $\mathbf{N}$ so that
$\theta_J = 0^\circ$ or $180^\circ$. For such cases, $\psi_J$ will
disappear from the waveform expression. These systems will be, on the
average, more luminous than edge-on systems and thus more likely to be
detected \cite{Canton:2014uja}.

For our purposes, the post-Newtonian (PN) formalism provides a
reasonable approximation to the observed gravitational waveform.
There are a variety of PN approximants available which differ in how
one deals with the energy, flux and balance equations (see
e.g. \cite{Buonanno:2009zt} for a recent review).  Our goal in this
paper is  not to study the differences between  various
approximants, but is rather to understand how precession affects the
size of the template bank. For this purpose, most approximants should
give similar results and our main consideration is computational
efficiency. For this purpose, since most of our computations are in
the frequency domain, it turns out to be very useful to work directly
with the Fourier transform of $h(t)$.  We shall use the frequency
domain model introduced in \cite{Lundgren:2013jla}.  An implementation
of this waveform model is publicly available in \cite{LSC_library},
where it is called the ``SpinTaylorF2'' model.

\subsection{Matched Filtering}
\label{sec:Matched Filtering}

Matched filtering is a methodology used to determine if data from a
gravitational wave detector $x(t)$, contains some signal of known
form, $h(t)$, or only Gaussian noise $n(t)$. Thus, in the absence of a
signal,
\begin{equation}
  x(t) = n(t) \,,
\end{equation}
and in the presence of a signal
\begin{equation}
  x(t) = h(t) + n(t) \,.
\end{equation}
If the noise is stationary, we can characterize it by the single-sided
power spectral density (PSD) $S_n(f)$ according to
\begin{equation}
  \langle \tilde{n}^\star(f) \tilde{n}(f^\prime)\rangle = \frac{1}{2} S_n(f)\delta(f-f^\prime)\,.
\end{equation}
Here the brackets $\langle\cdot\rangle$ denote an average over many
realizations of the noise, and $\tilde{n}(f)$ denotes the Fourier
transform of $n(t)$.

The PSD is used to define the inner product between two time-series $x(t)$ and $y(t)$:
\begin{equation}
  (x|y) := 4 \textrm{Re}\int_0^\infty \frac{\tilde{x}^\star(f)\tilde{y}(f)}{S_n(f)} df\,.
\end{equation}
This inner product is used to define the norm of a time series $x(t)$
and a normalized time series $\hat{x}$ in the usual way: 
\begin{equation}
  ||x|| := (x|x)^{1/2} \,,\qquad \hat{x} = x/||x||\,.
\end{equation}
The likelihood function $\Lambda$ can be shown to be \cite{Finn:1992wt,Jaranowski:1998qm}
\begin{equation}
  \log\Lambda = (x|h) - \frac{1}{2}(h|h)\,.
\end{equation}
The idealized procedure to search for a signal with unknown parameters
is to compute $\log\Lambda$ for all points (suitably
discretized) in a given parameter space and to find the point where $\log\Lambda$ is maximum.
The likelihood can be analytically maximized for certain parameters (such as the
initial phase $\phi_0$ and an overall constant amplitude) or by a Fast-Fourier
transform (such as the time of arrival $t_0$) (see
e.g. \cite{Allen:2005fk}), while other parameters (the so-called
intrinsic parameters) must be explicitly maximized over. These intrinsic parameters we denote as
$\lambda_i$. (We shall consider only binary systems with circular orbits and we
shall also not consider any parameters associated with the internal
structure of the neutron star.)

A template bank is a collection of waveforms $\{h_I\}$ labeled by the index
$I$.
Given a template bank, we would like to know how effective it is in
recovering a given signal $h$. This is quantified in terms of a number, namely the
\emph{fitting-factor} (FF) defined as, 
\begin{equation}
  FF(h, \{h_I\}) = \max_{I} \mu(h, h_I)\,,
\end{equation}
where
 \begin{equation}
  \mu(h,h_I) = \max_{t_0,\phi_0} (\hat{h}|\hat{h}_I(t_0,\phi_0))
 \end{equation}
is the \emph{match} between $h$ and $h_I$. $\mu(h,h_I)$ represents the fraction
of the optimal SNR of signal $h$ captured by the template $h_I$. The fitting
factor depends on a particular template bank and
a particular \emph{target} waveform $h$.  Since we will compute this
for a fixed template bank, we  usually drop its dependence on
$\{h_I\}$ and write $FF(h)$.

The loss in SNR can be quantified by the match between a signal and the 
nearest template and can be formulated geometrically \cite{Owen:1995tm,Owen:1998dk}.
The match between nearby points in parameter space can be approximated as
\begin{equation}
  \mu(\hat h(\lambda), \hat h(\lambda+d \lambda)) = 1 - g_{ij}d\lambda^i d\lambda^j  + \ldots
\end{equation}
with the metric
\begin{equation}
g_{ij} = -\frac{1}{2} \left.\frac{\partial^2 \mu\left(\hat h(\lambda), \hat h(\lambda')\right)}{\partial \lambda'_i \partial \lambda'_j}\right|_{\lambda'= \lambda}\,.
\end{equation}
This metric\footnote{For Gaussian stationary noise, one can show that
  the metric $g_{ij}$ is equivalent to constructing the scalar product
  $\frac{1}{2}\left.\left( \frac{\partial \hat h}{\partial \lambda_i}
    \right|\frac{\partial \hat h}{\partial \lambda_j}\right)$
  and projecting out the parameters $t_0$ and $\phi_0$.}  is useful in
quantifying the density of templates. The higher the metric determinant,
the higher the required template density for a fixed given allowed SNR
loss (which corresponds to a given minimal match).

For aligned waveforms, there is an analytic expression for the metric
\cite{Brown:2012qf,Harry:2013tca}.  However for the NSBH precessing
parameter space, we do not have an analytic way to calculate this
metric \cite{O'Shaughnessy:2015dsa}. We can approximate the precessing
metric by calculating the numerical derivative for small perturbations
in the waveform parameters ($\delta\lambda\rightarrow0$).  The numeric
metric is useful because it provides an independent validation of the
stochastic bank. For validation, we can qualitatively compare the
distribution of the determinant of the metric, and the distribution of
templates in the stochastic bank. As we shall see later, the density
of templates placed in the face-on bank will correlate with the
regions of the NSBH parameter space where the invariant volume
element, i.e. the square root of the determinant of $g_{ij}$ is high.

\subsection{Stochastic Placement}

\label{sec:Stochastic Placement}

The points in our stochastic template bank are populated using the
following steps.  The starting point for these could be either an
empty template bank or an existing ``seed'' template bank
\cite{Harry:2009ea,Babak:2008rb}:
\begin{enumerate}
\item Propose a physically viable point in parameter space $p$
  following some probability distribution (we call this
  distribution the \emph{proposal distribution}).  If we are starting
  with an empty bank, then the first proposed point will always be
  accepted.
\item Calculate the match of the waveform at $p$ with all the
  waveforms previously accepted into the bank.
\item Append the candidate to the bank if all the matches are below some
  threshold, known as the \emph{minimal match}. We shall take the
  minimal match to be 97\%.
\item Repeat the previous steps until a convergence condition is
  achieved. We shall take the convergence criteria to be: continue the
  process until, in the previous 1000 trials, only 30 or fewer points
  have been accepted.
\end{enumerate}

The resulting template bank will, of course, depend on the proposal
distribution that we start with. If we had sufficiently reliable
astrophysical information on spin orientation, mass distributions
etc., we could tailor the template bank appropriately.  In the absence
of any such prior information, we need to apply some other criteria
for the proposal distribution.  A well motivated choice is to choose
the distribution according to the value of the determinant of the
metric $g_{ij}$ (this is the choice made in
e.g. \cite{Messenger:2008ta}); indeed a geometric placement algorithm
would satisfy this condition.  However, this is not the only
possibility and we shall discuss our choice below.

\subsection{Proposal Distribution}
\label{sec:proposalDist}

We shall start with the assumption that the binary system is face-on,
i.e. $\mathbf{J}$ is pointing either directly towards or directly away from the
detector and we shall fix the sky-location to be directly overhead the
detector.  With these assumptions, we are left with a five dimensional
problem: the two masses $M_1$ and $M_2$, and the three components of
the black-hole spin $\mathbf{S}$.

Even with these assumptions, the full problem is a significant
computational challenge.  An important issue is that the stochastic
placement algorithm is not easy to parallelize.  Imagine trying to
divide the full parameter space into smaller sub-regions and applying
the procedure outlined above to each of these sub-regions.  Note that
in the second step of the procedure outlined in the previous section,
we need to check the match of a new waveform with \emph{all} of the
previously accepted waveforms in the template bank.  Thus, in
principle, each sub-region needs to be aware of the points that have
been accepted in the other sub-regions.  Dealing with each sub-region
independently could lead to a significant over-coverage,
i.e. accepting many more points than necessary.  

If we could find sub-regions which are uncorrelated from each other
(by a suitable choice of coordinates) and if the sub-regions were
sufficiently large, then the parallelization would be close to
optimal.  While we do not have the optimal coordinates for this
purpose, it turns out that a simplified version of the so-called \emph{chirp times}
($\tau_{0},\tau_{3}$) \cite{Sengupta:2003wk, Canton:2014ena} are a good approximation:
\begin{eqnarray}
\tau_{0}&=&\mathcal{M_{\mathrm{C}}}^{-5/3}\label{eq:Tau0}\\
\tau_{3}&=&\mathcal{M_{\mathrm{C}}}^{-2/3}(\nu(1-\nu))^{-3/5}(4\pi-\beta_C)\label{eq:Tau3}
\end{eqnarray}
where
\begin{equation}
  \beta_C=\frac{1}{12}(38\nu^2+75\nu) \chi^\parallel\label{eq:Beta_C}\,.
\end{equation}
The chirp time was first introduced in \cite{Sathyaprakash:1991mt}
 as the time taken
for the Gravitational Wave (GW) signal to reach coalescence starting from some
initial frequency.  Chirp times are also the coordinates
typically used in geometric methods for template placement
\cite{Cokelaer:2007kx} and is also the coordinate where the parameter
space metric for binary inspiral systems is most easily understood
(see e.g. \cite{Keppel:2013kia}).

We wish to cover the $(\tau_0,\tau_3)$ space uniformly. In particular,
while constructing the template bank for a particular rectangular
region, we would like to ensure that we generate templates only for
that rectangular region.  If we were to pick values of
$M_1, M_2, \mathbf{S}$ directly, this would not be guaranteed.  We
therefore follow the following steps:
\begin{enumerate}
\item Generate values of $\tau_0$ and $\tau_3$ randomly within the
  rectangular region under consideration following a uniform
  distribution.
\item The value of $\tau_0$ determines the chirp mass $\cal{M}_C$, but
  $\tau_3$ depends on both $\nu$ and $\chi^\parallel$.  Our strategy is
  to then pick a value of $q$ which, along with the chosen value of
  $\tau_3$ determines $\chi^\parallel$.  The value of $\nu$ is chosen
  randomly assuming that $M_1$ and $M_2$ are uniformly distributed in
  their allowed ranges.  In practice, we draw random values of $M_1$
  and $M_2$ from uniform distributions, which determines $\nu$. Given
  $q$ and $\tau_3$, we solve Eq.~(\ref{eq:Tau3}) for $\chi^\parallel$.
\item To fix the component of $\mathbf{S}$ perpendicular to
  $\widehat{\mathbf{L}}$, we note that the total spin magnitude $\chi$
  is bounded below by $\chi^\parallel$.  We pick a value of
  $\chi^\perp$, such that $||\chi||$ is uniformly distributed between $\chi^\parallel$ and $1$.
\item Finally, we choose $\alpha_0$ uniformly between 0 and $2\pi$.  
\end{enumerate}
This procedure ensures that the proposal distribution covers all
possible precessing binary configurations.  Lower values of $\tau_{3}$
get mapped to the more aligned systems, i.e. larger values of
$\chi^\parallel$, while lower values of $\tau_{0}$ are mapped to
systems with higher total mass, $M$.  While the resulting distribution
of points in the physical parameters
$(M_1, M_2, \chi^\parallel, \chi^\perp, \alpha_0)$ will not be
completely physical, our final results are not very sensitive to this choice
of distribution.

\section{The precessing face-on template bank [FOB]}
\label{sec:bank}

The total range of chirp times corresponding to our parameter space is
broken up into $938$ smaller ``chirp time boxes''.  A stochastic
template bank is constructed for each box independently and the $938$
template banks are then concatenated.  Parallelizing the stochastic
placement is computationally advantageous, because the stochastic
placement algorithm's efficiency scales with the square of the number of
templates placed \cite{Brown:2012qf,Harry:2013tca}. By splitting up
these regions, we limit the number of comparisons needed for each
stochastic template bank candidate to decide whether it should be accepted or
not. However, speeding up the algorithm comes at the
cost of over-coverage between neighboring boxes which we shall discuss
towards the end of this section.

\begin{figure}
 \includegraphics[scale=0.47]{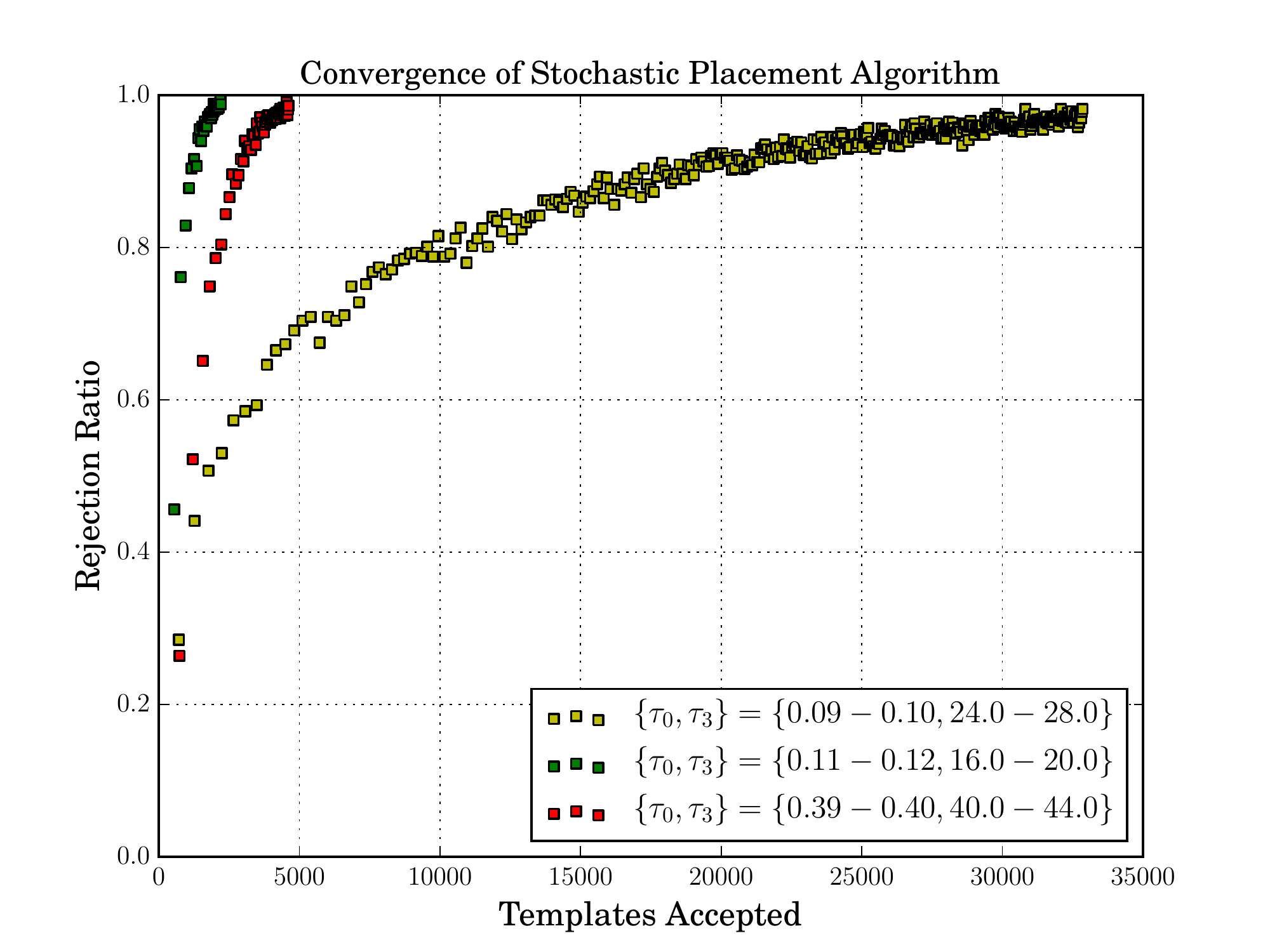}
 \caption{Convergence curves of three different boxes used to construct the FOB.}
 \label{fig:Convergence} 
\end{figure}

The parameter space metric discussed earlier also plays a role in
reducing the computational cost, and in particular we use the metric
for the space of \emph{aligned-spin} waveforms \cite{Harry:2013tca}.
The goal is to minimize the number of times that we need to calculate
the match.  For a proposed parameter space point, we consider only
those waveforms which have a match of better than $70\%$ with the
proposed waveform as computed by the aligned spin metric. The full
match is computed only for the waveforms in the template bank which
cross this threshold.  The $70\%$ threshold was found by trial and
error and is low enough that we do not miss any templates close to the
proposed waveform. Fig.~\ref{fig:Convergence} shows the convergence of
the match for three particular boxes in $(\tau_0,\tau_3)$ space.

\begin{figure}
\includegraphics[scale=0.47]{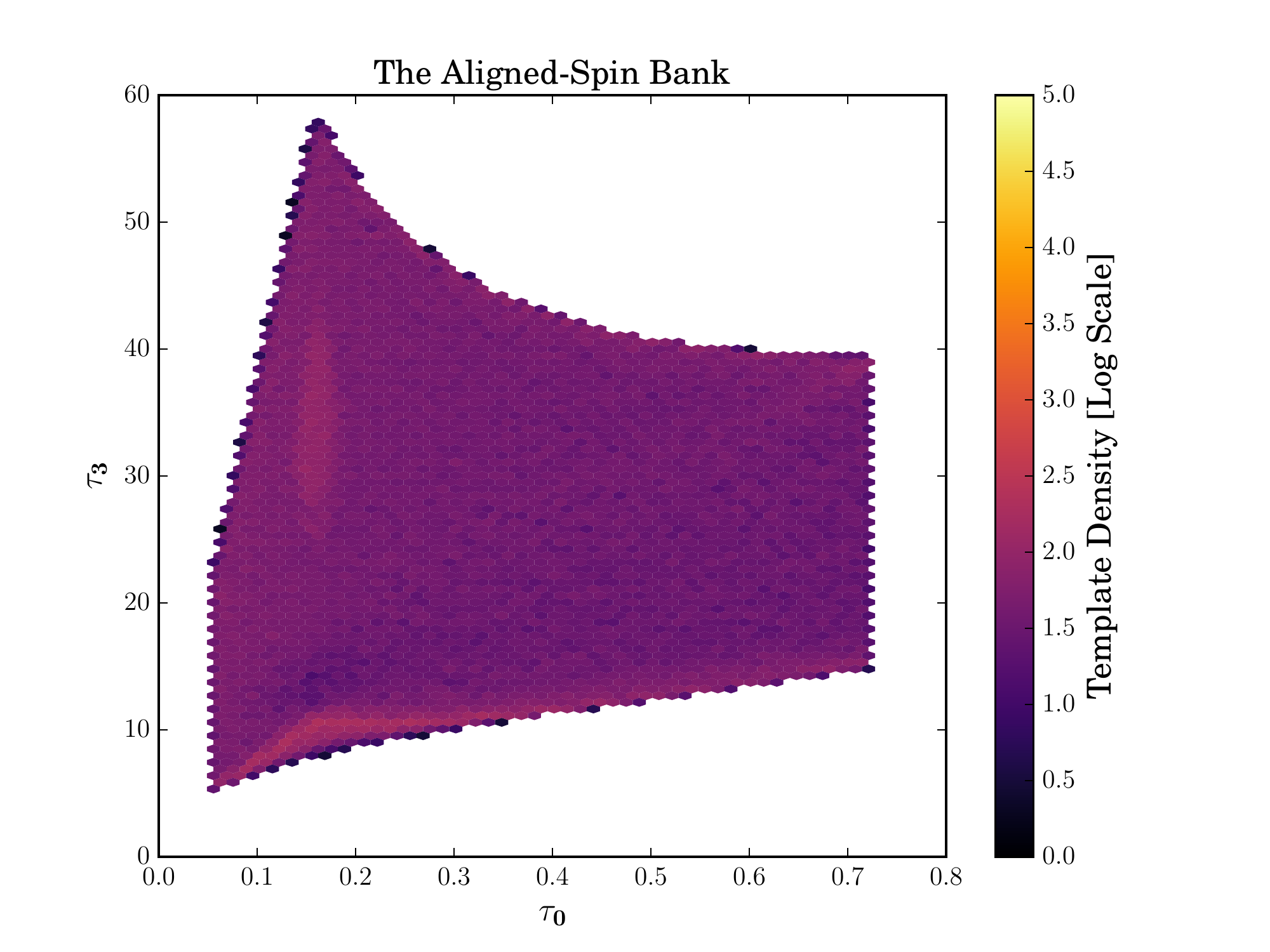}
\caption{The aligned spin bank (ASB) in chirp time coordinates. The color bar
density scale is the same as in Figure \ref{fig:FaceOnBank} for ease of
comparison. Each hexbin has dimensions $\{\Delta\tau_0=0.014,\Delta\tau_3=1.0\}$.}
\label{fig:TitosBank}
\end{figure}

\begin{figure}
\includegraphics[scale=0.47]{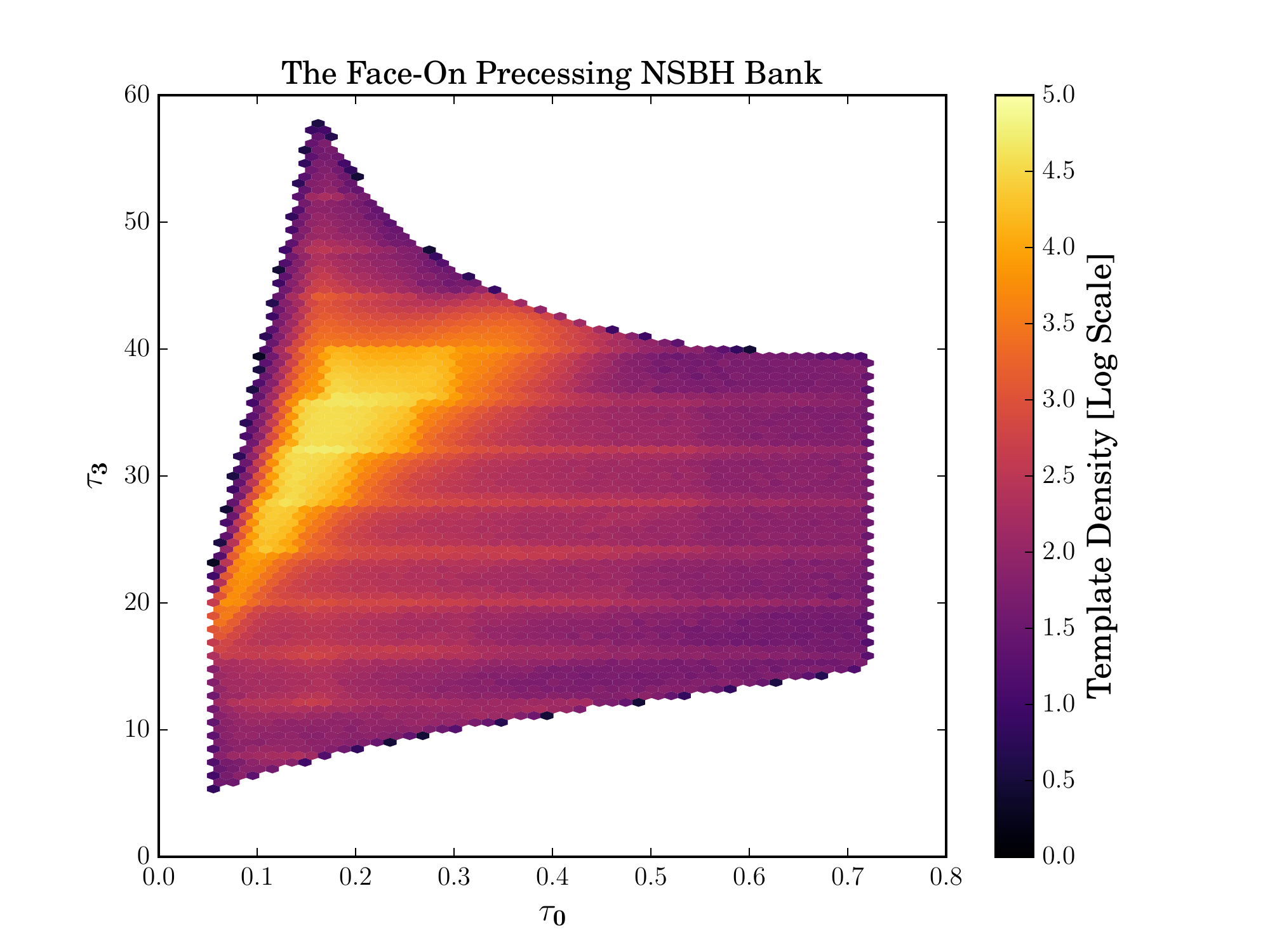} 
\caption{The FOB in chirp time coordinates. Each hexbin has dimensions $\{\Delta\tau_0=0.014,\Delta\tau_3=1.0\}$.}
\label{fig:FaceOnBank}
\end{figure}

\begin{figure}
\includegraphics[scale=0.47]{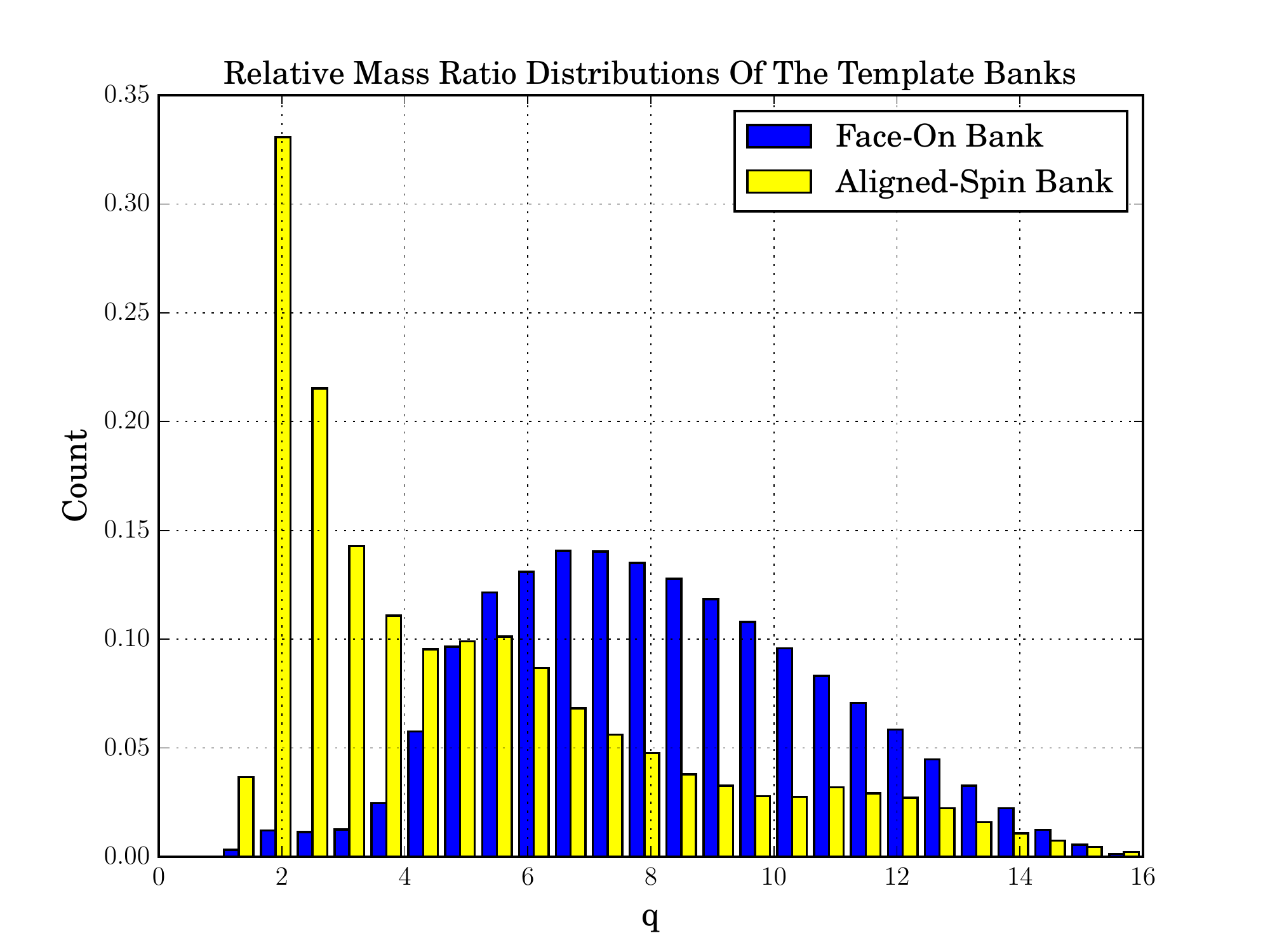} 
\caption{Normalized distribution of the mass-ratio $q$ for the face-on precessing
  template and ASB.}
\label{fig:FaceOnBankMassRatio}
\end{figure}

Before presenting the result of the above procedure and discussing some
properties of the precessing face on template bank (FOB), we briefly describe an
aligned-spin bank (ASB) which we shall use as a reference for comparison. Such
bank covers the same space of masses and aligned spin components, but ignores
precession. It is constructed via stochastic placement using non-precessing,
inspiral-only post-Newtonian templates (namely the ``TaylorF2'' model in
LALSimulation \cite{LSC_library}) and contains $130,646$ templates. The template
density is shown in Figure~\ref{fig:TitosBank} in chirp time coordinates
$(\tau_0,\tau_3)$ and in such coordinates it is approximately constant.
The ability of a similar bank at detecting aligned-spin and precessing NSBH
systems has been characterized in previous studies \cite{Canton:2014ena,
Canton:2014uja}.
In contrast, the template bank for precessing face-on systems is shown
in Figure~\ref{fig:FaceOnBank}. It contains $6,908,681$ templates -- a
dramatic increase compared to the ASB.  The densest parts
are in the {\it high mass ratio, and highly anti-aligned spin ($\kappa<-0.5$) region}
of the bank. More than half of the total number of templates are placed in this region.
Figure~\ref{fig:FaceOnBankMassRatio} shows the distribution of the mass
ratio in the FOB and ASB, thereby demonstrating that the vast majority
of points in the FOB consist of asymmetric systems (with mass-ratio
$q>4$) in contrast to the ASB which is dominated by more symmetric
systems.

\begin{figure}
\includegraphics[scale=0.47]{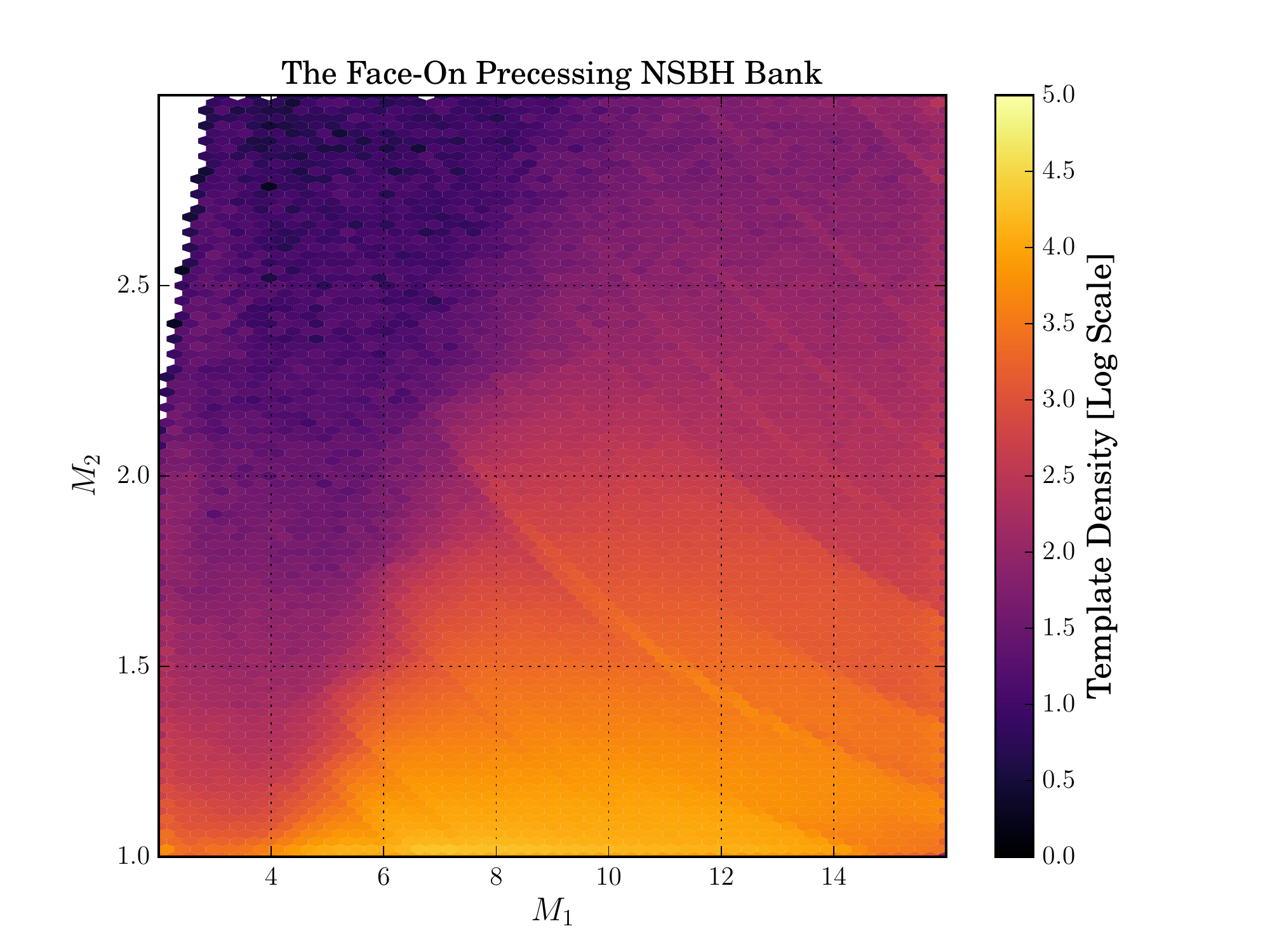} 
\caption{The FOB in solar mass ($M_1,M_2$) coordinates. As
  before, the color bar is scaled with respect to the density of
  templates per bin $\{\Delta M_1 =0.28,\Delta M_2=0.02\}$ . }
\label{fig:FaceOnBankM1M2}
\end{figure}

\begin{figure}
\includegraphics[scale=0.47]{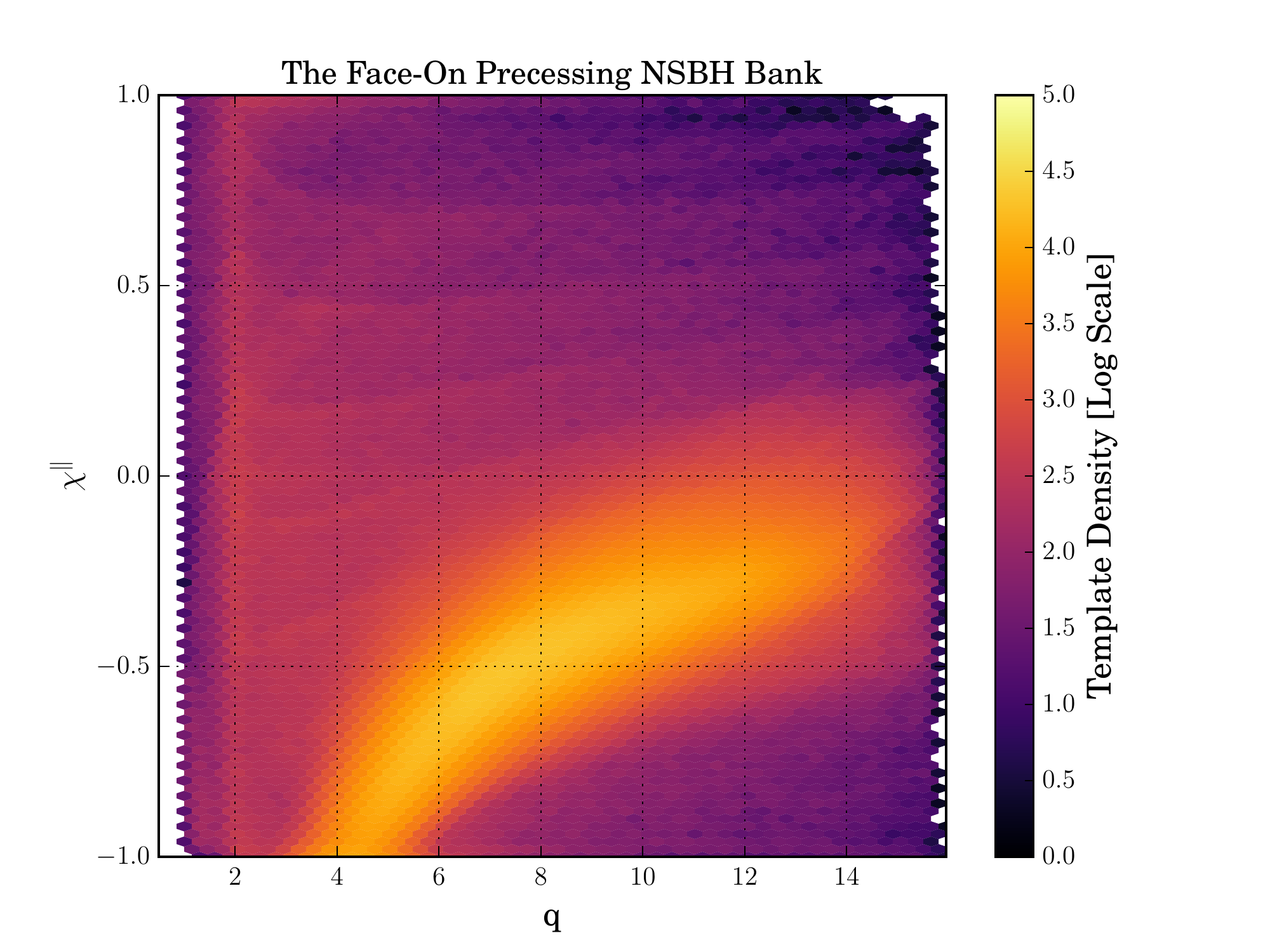} 
\caption{The FOB in ($q,\chi^\parallel$) coordinates. As
  before, the color bar is scaled with respect to the density of
  templates per hexbin $\{\Delta q =0.3,\Delta\chi^\parallel=0.04\}$ . }
\label{fig:FaceOnBankqS1L}
\end{figure}

\begin{figure}
\includegraphics[scale=0.75]{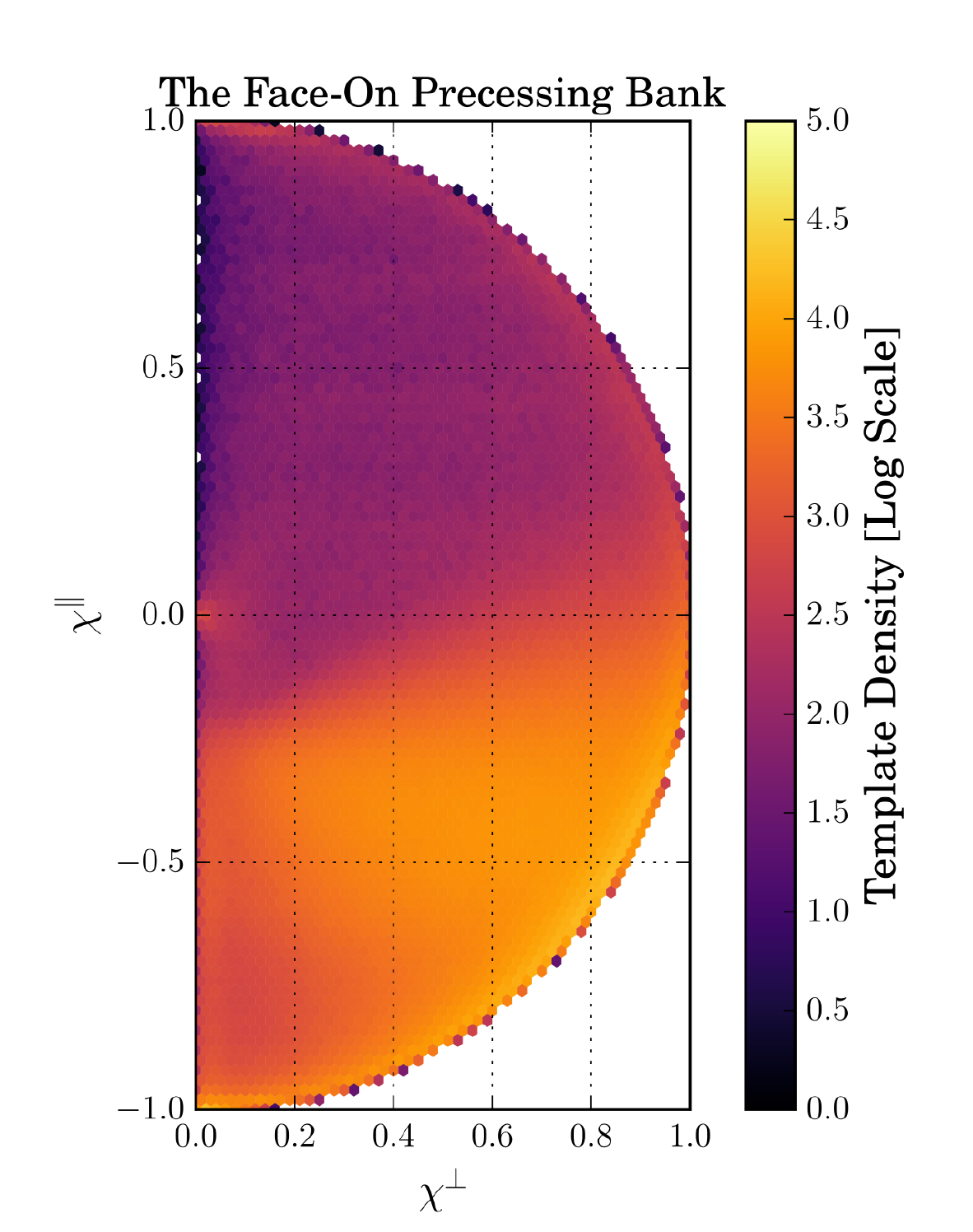}
\caption{The spin distribution of the FOB.  The
  y-axis is the component of spin parallel to the orbital angular
  momentum $\mathbf{L}$ and the x-axis is the component of the spin
  perpendicular to $\mathbf{L}$. Each hexbin has dimensions $\{\Delta\chi^{\parallel}=0.04,\Delta\chi^{\perp}=0.02\}$.}
 \label{fig:FaceOnBankspin}
\end{figure}

\begin{figure}
\includegraphics[scale=0.47]{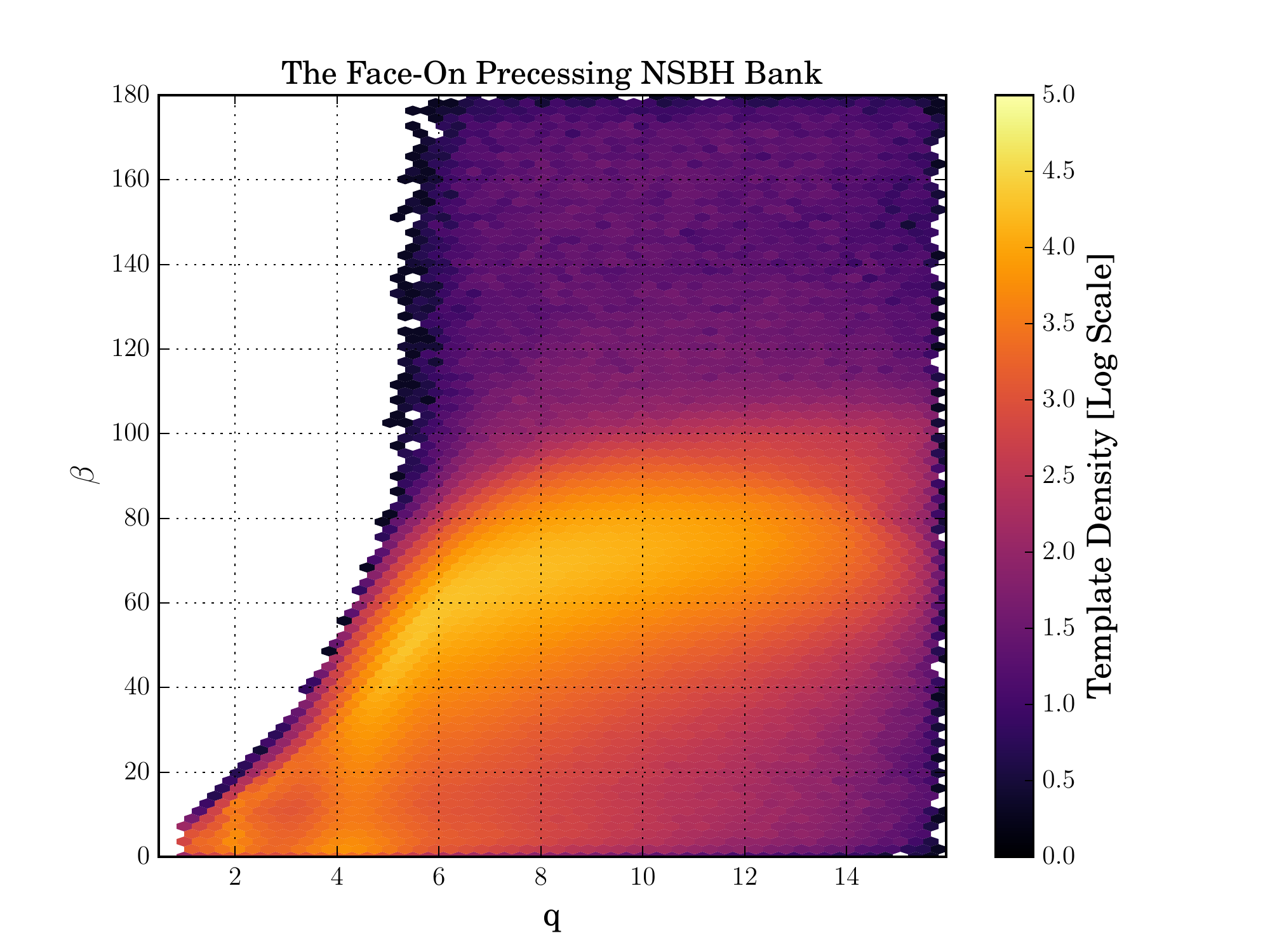}
\caption{The $\beta\,$ and mass ratio, $q$, distribution of the FOB. Each hexbin has dimensions $\{\Delta q=0.3,\Delta\beta\,=4^{\circ}\}$.}
\label{fig:FaceOnBeta}
\end{figure}

Figures~\ref{fig:FaceOnBankM1M2}, ~\ref{fig:FaceOnBankqS1L},
\ref{fig:FaceOnBankspin} and
\ref{fig:FaceOnBeta} display the precessing template bank in different
slices of the parameter space.  Figure~\ref{fig:FaceOnBankqS1L} shows
the template bank density in the $(q,\chi^\parallel)$
plane. Figure~\ref{fig:FaceOnBankspin} shows the distribution of
templates in the $(\chi^\perp, \chi^\parallel)$ plane. Finally
Figure~\ref{fig:FaceOnBeta} gives the template bank distribution in
the $(q,\beta \, )$ plane. In Figure~\ref{fig:FaceOnBankspin}, we note that,
higher template densities occur in the higher values of spin-orbit
misalignment, which in-turn indicates higher precession.

As a result of breaking up the parameter space into independent boxes,
it is to be expected that the algorithm will place more templates than
necessary at the borders between adjacent boxes.  This creates
so-called ``gridlines'' in the bank which are clearly visible in
Figure~\ref{fig:FaceOnBank}. These are an artifact of splitting up
the parameter space into independent regions.  It results in having a
larger number of templates than necessary.  However, we shall see that
this is not a large effect for the chirp time boxes that we have
chosen.

The gridlines were most pronounced at the edges of the boxes along the
vertical direction, which implies that there is a degeneracy along the
$\tau_{3}$ direction.  The gridlines along the $\tau_0$ direction are
much less pronounced.  This is not surprising since $\tau_0$ is
determined entirely by the chirp mass $\mathcal{M}_C$, and it is well
known that $\mathcal{M}_C$ is the parameter best determined from the
inspiral phase \cite{Ohme:2013nsa}. 

\begin{figure}
\includegraphics[scale=0.47]{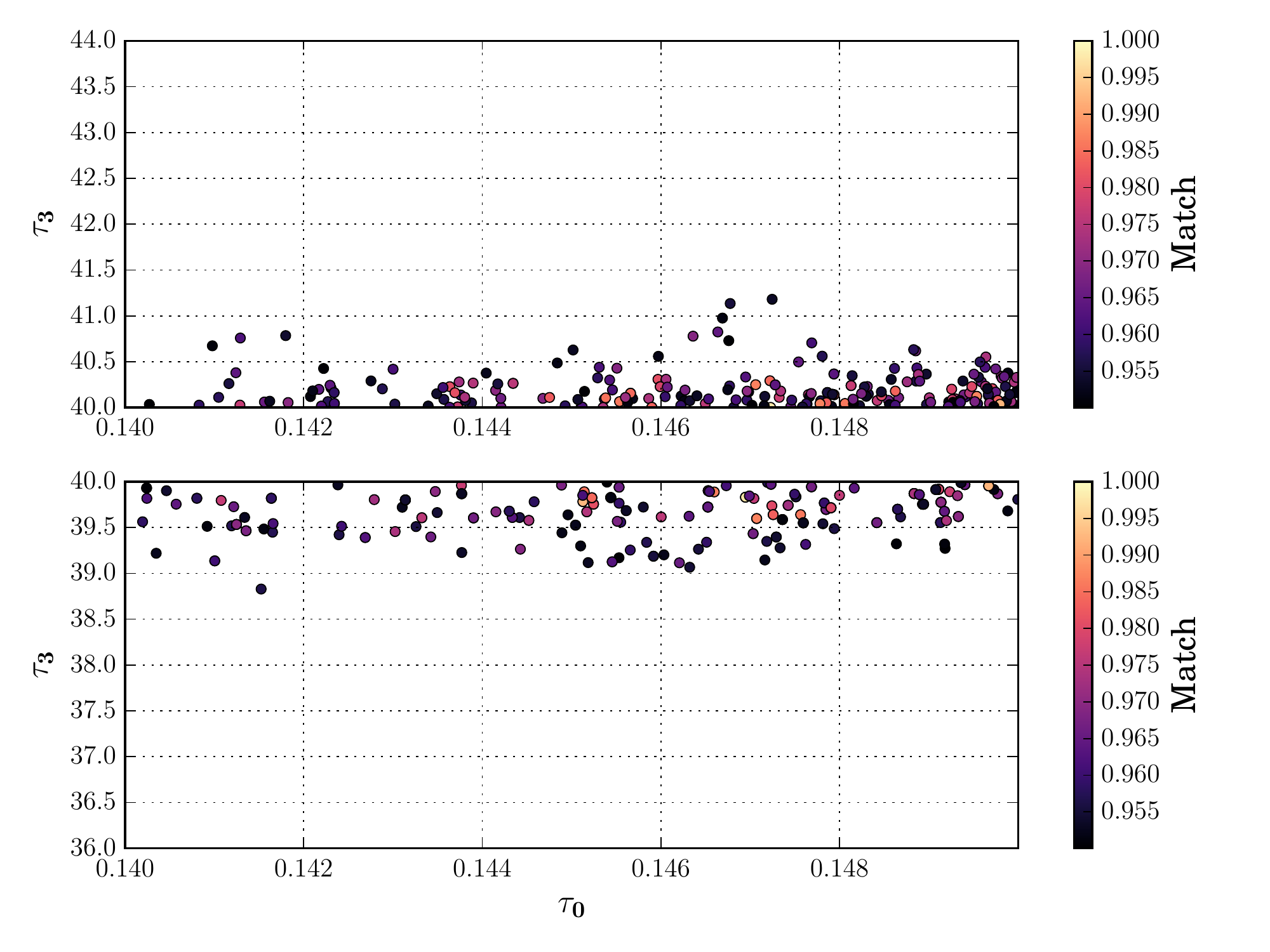}
\protect\protect\protect\caption{Plot of the templates in the box $(\tau_{0},\tau_{3})=(0.14-0.15,40-44)$
that had an overlap greater than $95\%$ with templates placed in
the box, $(\tau_{0},\tau_{3})=(0.14-0.15,36-40)$.}
\label{fig:vertical_gridline_comp}
\end{figure}

This suggests also that it should be possible to
replace $\tau_3$ by a better coordinate leading to fewer correlations.
Regardless, we shall now quantify the correlations between adjacent
boxes in the $\tau_3$ direction.  Also, in order to optimize the size
of the chirp time boxes, it was crucial to estimate how far these
gridlines overlapped into adjacent boxes. To study this issue, we
looked at two adjacent boxes in $\tau_{3}$.  By taking points in the
lower box and calculating the overlap with every point in the above
box, we determined the extent of the overcoverage.
Fig.~\ref{fig:vertical_gridline_comp} displays the templates in
adjacent boxes which have an overlap greater than 95\% with templates
in the adjacent box.  The extent of these templates extends to about
25\% of the box in the $\tau_3$ direction. However, the number of such
templates is only about 7\% of the total number of templates in the
upper box, and 1\% for the lower box. 

\begin{figure}
\includegraphics[scale=0.75]{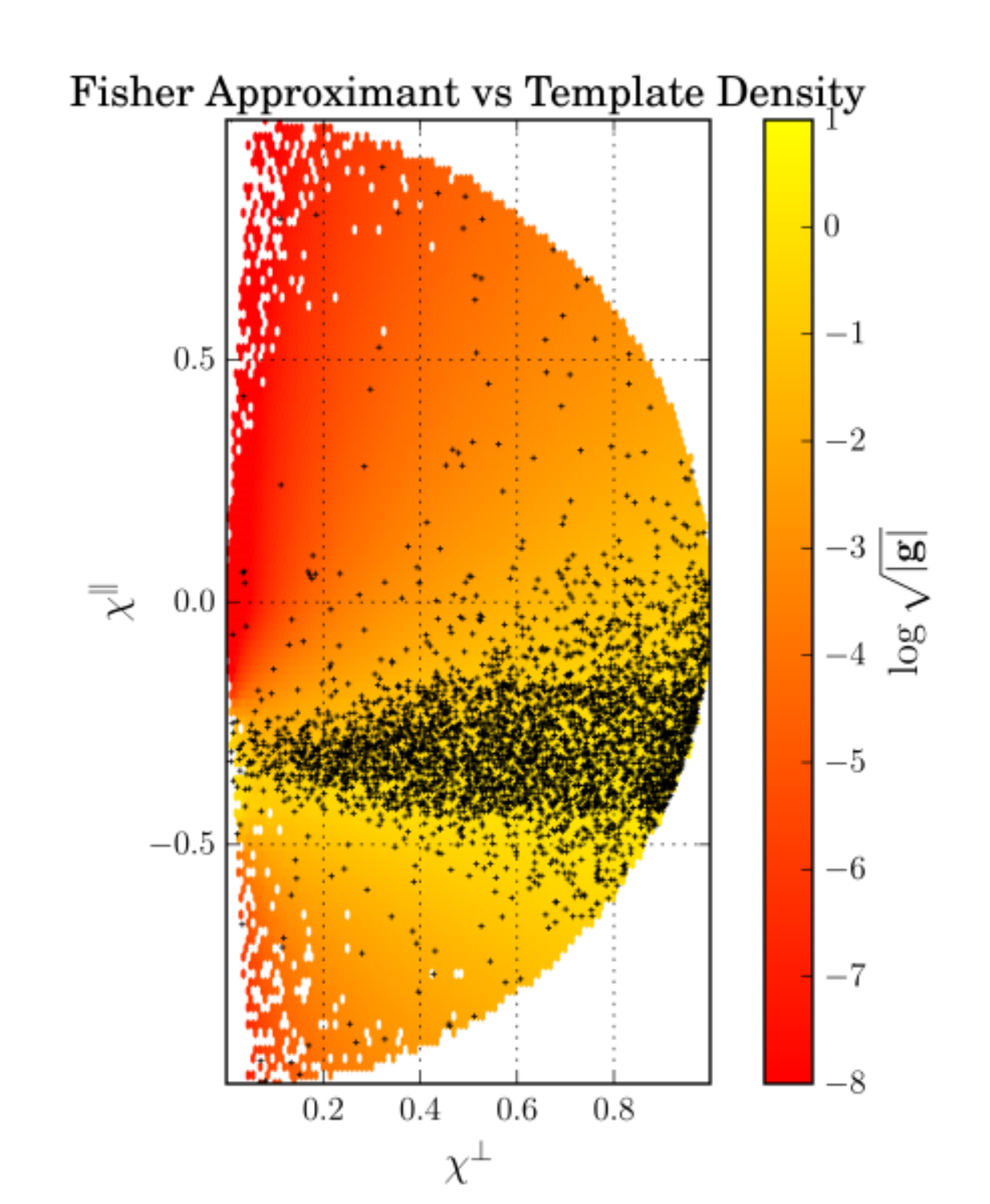}
\caption{Comparison of the stochastic bank and the metric approximant $\mathbf{\chi}$ distributions for fixed masses $\{15M_\odot,1.4M_\odot\}$. The color map represents  $\log \sqrt{|g|}$ and the scattered points denote the templates in the stochastic bank with masses within 1\% of $\{15M_\odot,1.4M_\odot\}$. Each hexbin has dimensions $\{\Delta\chi^{\parallel}=0.04,\Delta\chi^{\perp}=0.02\}$.}
\label{fig:facOn_15_1p4_Scmpts}
\end{figure}

To conclude this section, we validate the distributions obtained above
by a numerical calculation of the Fisher matrix.  If one were able to
carry out a geometric template placement procedure, the density of
templates would be proportional to the invariant volume element, i.e. to
the square root of the determinant of $g_{ij}$.  The same is generally
true for probabilistic methods of template placement.  We compute
$g_{ij}$ and its determinant directly by numerically computing the
overlap between the derivatives of neighboring waveforms and compare this with the actual
distribution of templates obtained in the template bank.
Figure~\ref{fig:facOn_15_1p4_Scmpts} shows the contour plot of
$\log\sqrt{\mathrm{| g |}}$ for the $\{15M_\odot, 1.4M_\odot\}$
case. Also shown are the points in the template bank whose masses are
within 1\% of these mass values demonstrating qualitative agreement
between the two entirely different calculations.  Similar results are
obtained for other values of the masses and other slices of the
parameter space.  This agreement between the two independent
calculations provides a sanity check and indicates that the great
increase in the number of templates is a real feature of the space of
precessing
waveforms. Using a different detection statistic as in
\cite{Harry:2016ijz} helps ameliorate the problem somewhat, but does
not eliminate it.

\section{Effectualness of the template bank}
\label{sec:effectualness} 

In this section we estimate the effectualness of the FOB for different
populations of NSBH systems and compare it with the ASB. When calculating
matches, all simulations use a lower frequency cutoff of $30$ Hz and an upper
cutoff of $4400 / (M_1+M_2)$ Hz, which is the frequency corresponding to the
innermost stable circular orbit of a Schwarzschild black-hole with mass equal
to $M_1+M_2$.

\begin{figure}
\includegraphics[scale=0.47]{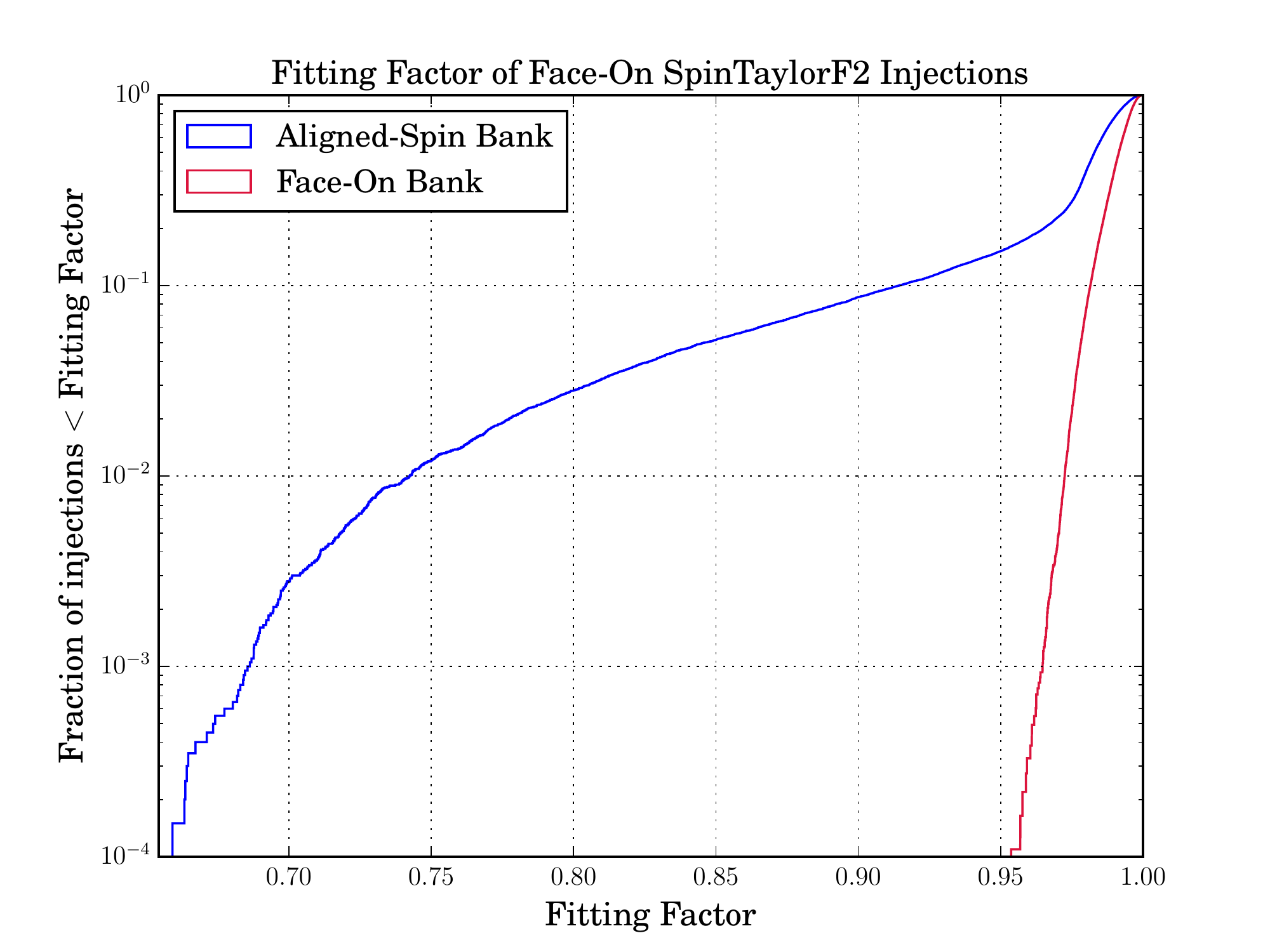} 
\caption{Cumulative histogram showing the recovered fitting factor of the
  face-on-precessing and aligned spin template banks for face-on
  SpinTaylorF2 injections.}
\label{fig:faceon_banksim_match_FOB_vs_ASB_F2}
\end{figure}

\begin{figure}
\includegraphics[scale=0.47]{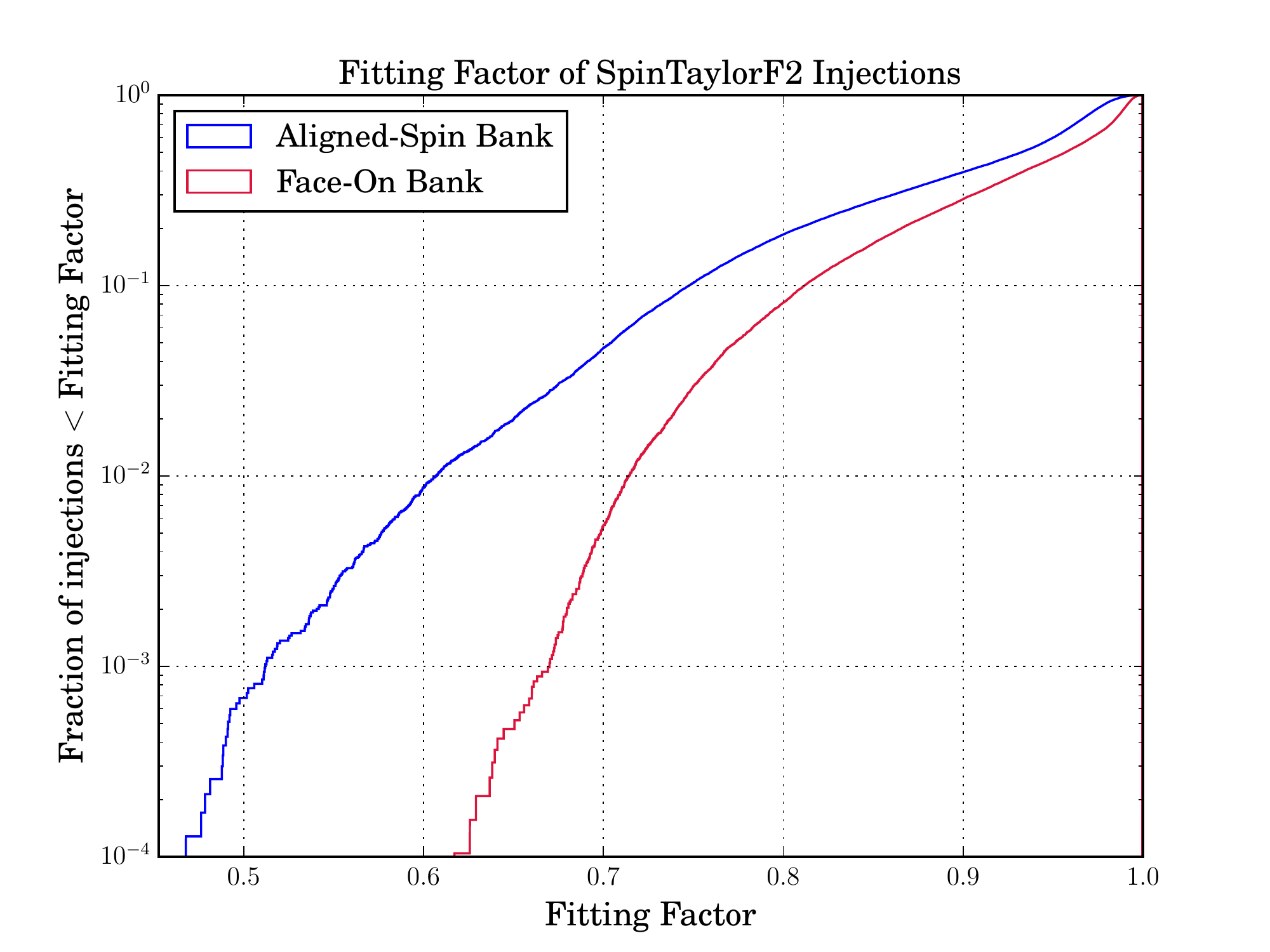} 
\caption{Cumulative histogram showing the recovered fitting factor of the
  face-on-precessing and aligned spin template banks for
  SpinTaylorF2 injections with the component masses distributed uniformly within their respective ranges, spins distributed uniformly in $\kappa$, and $\widehat{\mathbf{J}}$ distributed uniformly over the sphere.}
\label{fig:all_banksim_match_FOB_vs_ASB_F2}
\end{figure}

First, we want to consider the waveforms which were used to construct
the FOB, namely the SpinTaylorF2 waveforms and we want to
compare the ASB with the FOB for
precessing waveforms.  In order for a bank to recover signals
effectively, it must be able to recover NSBH systems over a range of
mass and spin values and orientations of $\mathbf{\hat{J}}$.  We
consider two cases: i) by constraining the injections to be face-on
NSBH systems, we look at how well the FOB and ASB could recover
SpinTaylorF2 injections from the same proposal distribution used to
construct the FOB, and ii) for arbitrary orientations of the total
angular momentum (i.e.  $0^\circ < \theta_J < 180^\circ$). In both
cases, we considered injections over the same $\{M_1,M_2\}$ parameter
space as before, i.e.  $2M_\odot < M_1 < 16M_\odot$ and
$1 M_\odot < M_2 < 3M_\odot$.
Figs.~\ref{fig:faceon_banksim_match_FOB_vs_ASB_F2} and
\ref{fig:all_banksim_match_FOB_vs_ASB_F2} show the recovered fitting
factors for the ASB and FOB banks for these two cases.
Figure~\ref{fig:faceon_banksim_match_FOB_vs_ASB_F2} shows the case when
the injections are face-on. This is what the FOB was built for and
indeed, the plot shows that the FOB greatly outperforms the ASB.  The
fitting factors are worse than 97\% for no more than 1\% of the
injections.
Figure~\ref{fig:all_banksim_match_FOB_vs_ASB_F2} shows the corresponding
result when the injections are not constrained to be face-on. The
recovered matches are reduced, but the FOB still outperforms the ASB
over the full mass range.

\begin{figure}[H]
\includegraphics[scale=0.47]{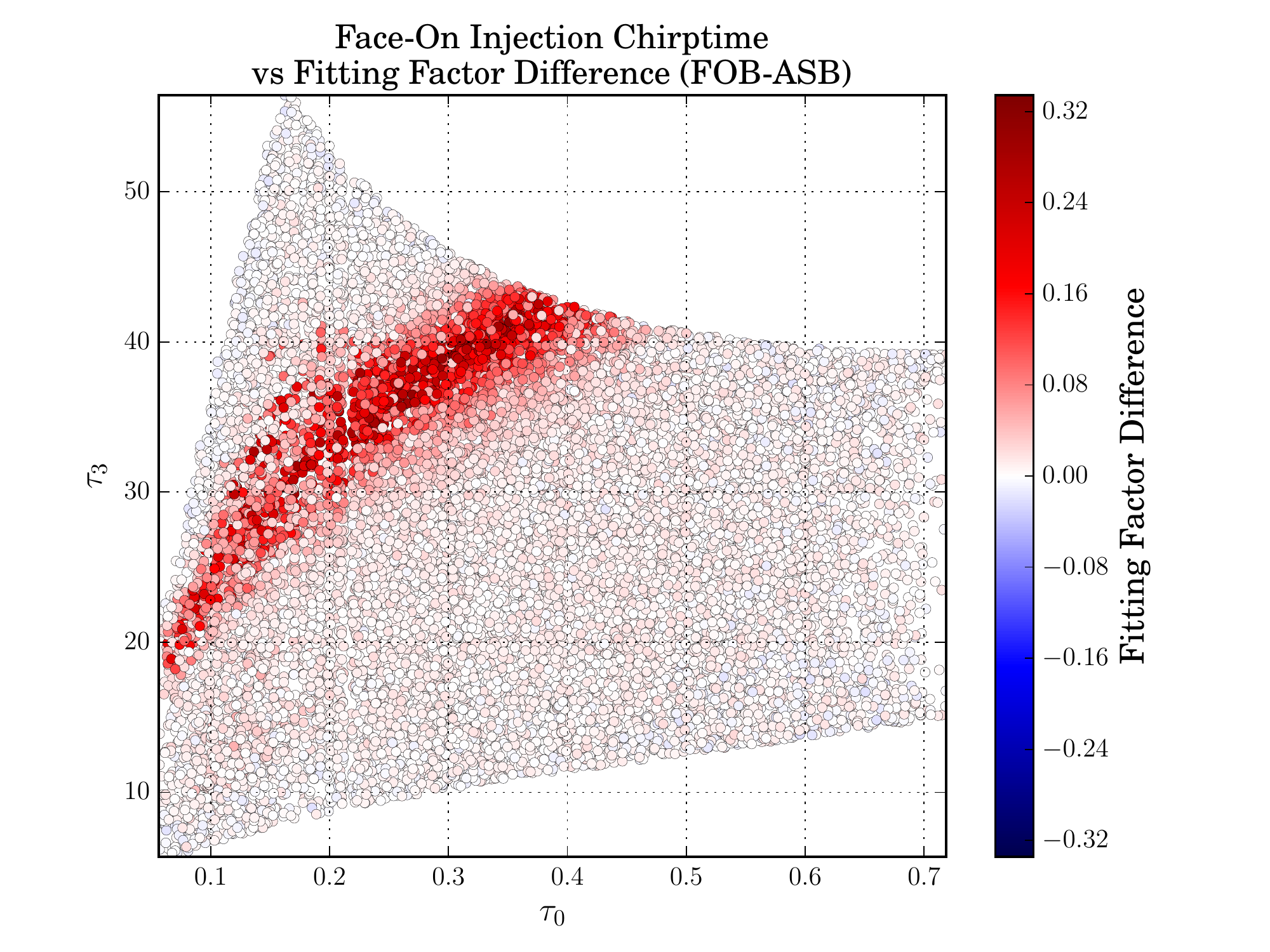} 
\caption{A plot of the difference in the recovered fitting factor
  between the precessing and aligned template banks over the
  $\{\tau_0, \tau_3\}$ parameter space for face-on SpinTaylorF2
  injections. }
\label{fig:matchdiff_tau0tau3_faceon}
\end{figure}

\begin{figure}
\includegraphics[scale=0.47]{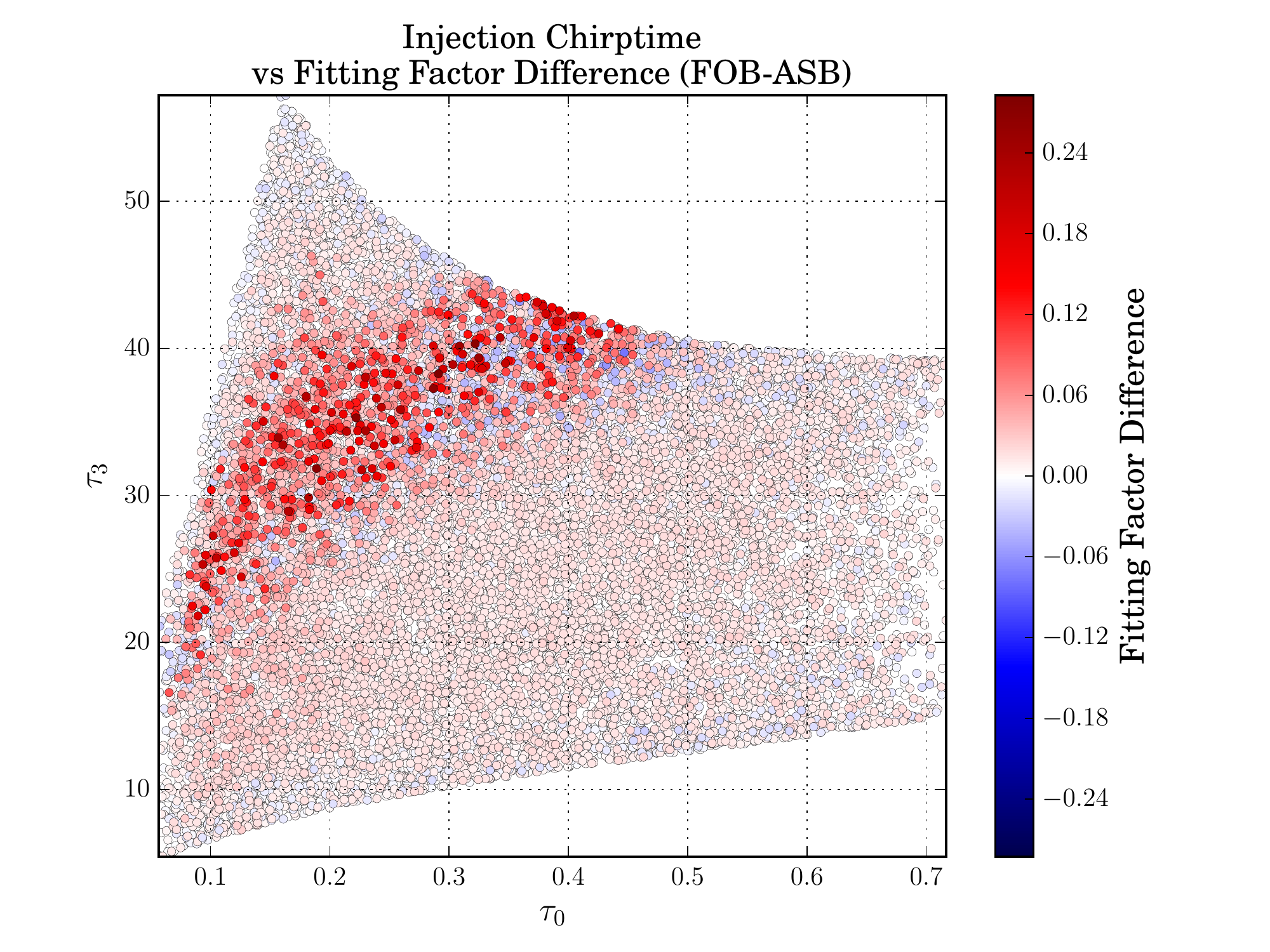} 
\caption{A plot of the difference in the recovered fitting factor
  between the precessing and aligned template banks over the
  $\{\tau_0, \tau_3\}$ parameter space for SpinTaylorF2 injections 
  that are distributed uniformly in chirp time, $\{\tau_0,\tau_3\}$,
  with $\widehat{\mathbf{J}}$ distributed uniformly over the sphere.}
\label{fig:matchdiff_tau0tau3}
\end{figure}

\begin{figure}
\includegraphics[scale=0.47]{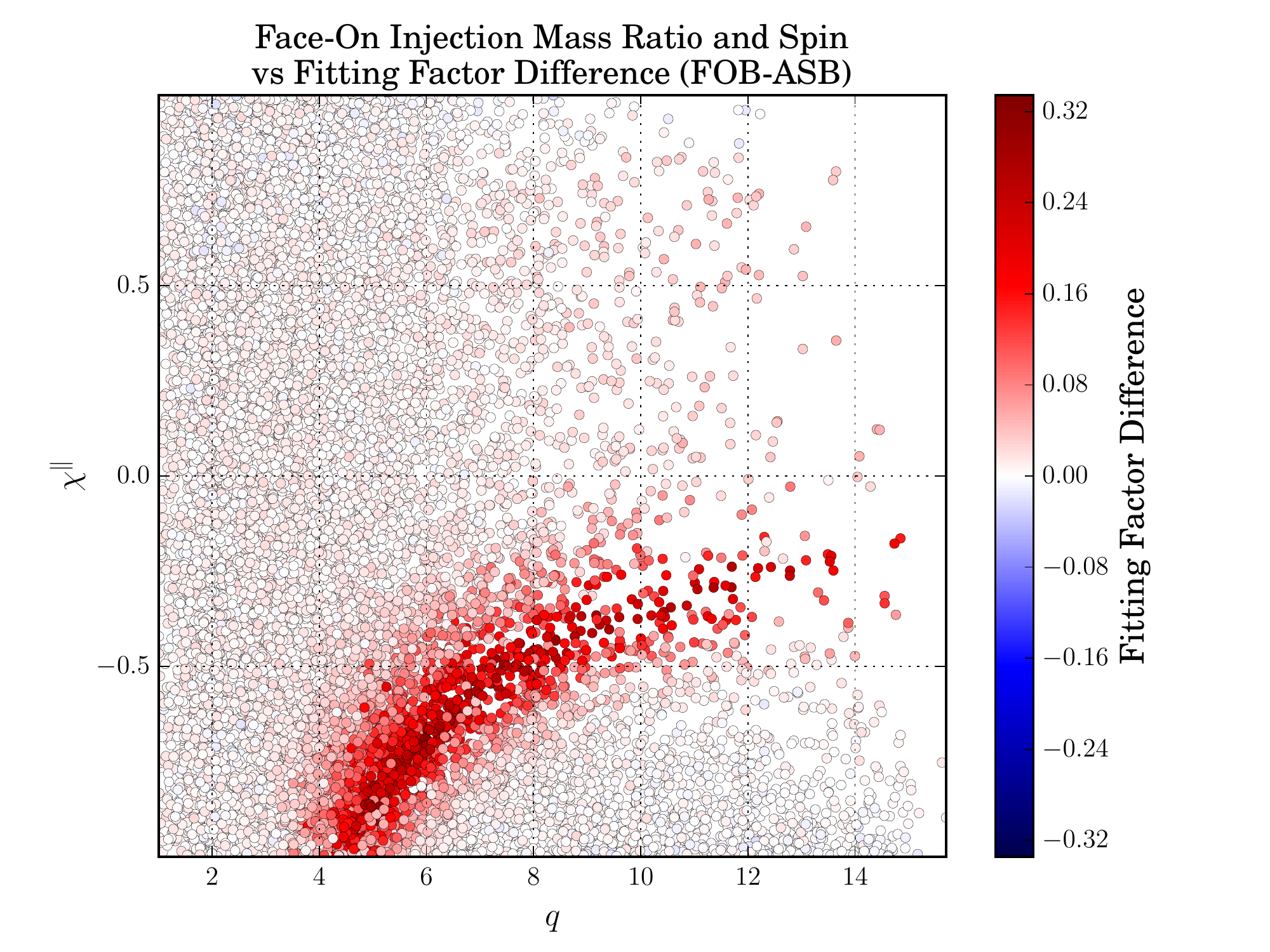} 
\caption{A plot of the difference in the recovered fitting factor
  between the precessing and aligned template banks over the
  $\{q, \chi^\parallel\}$ parameter space for
  face-on SpinTaylorF2 injections. }
\label{fig:matchdiff_qk_faceon}
\end{figure}

\begin{figure}[H]
\includegraphics[scale=0.47]{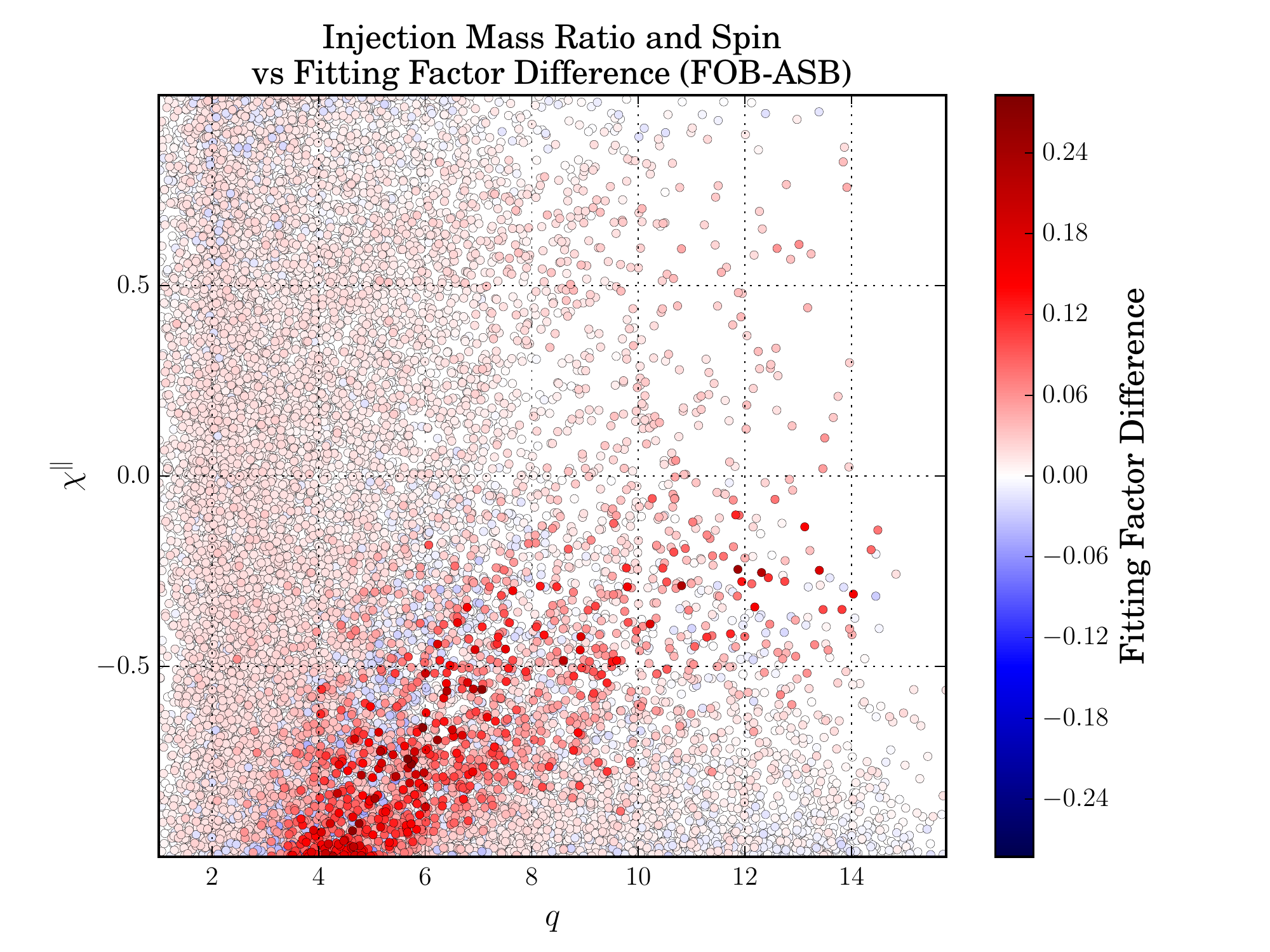} 
\caption{A plot of the difference in the recovered fitting factor
  between the precessing and aligned template banks over the
  $\{q, \chi^\parallel\}$ parameter space for SpinTaylorF2 injections
  that are distributed uniformly in chirp time, $\{\tau_0,\tau_3\}$,
  with $\widehat{\mathbf{J}}$ distributed uniformly over the sphere.}
\label{fig:matchdiff_qk}
\end{figure}

\begin{figure}
\includegraphics[scale=0.47]{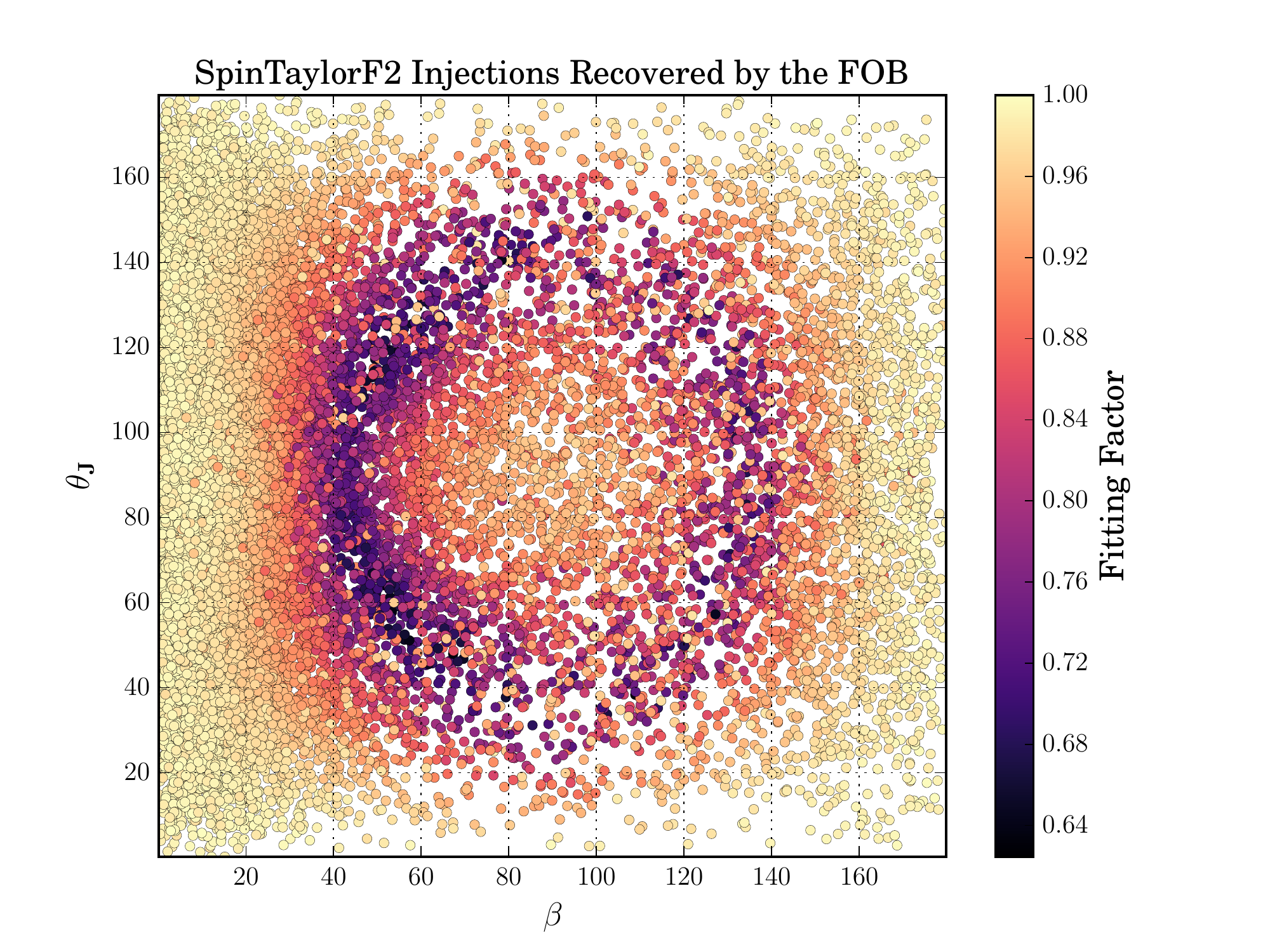} 
\caption{Recovered fitting factor of the precessing template bank over
  the $\{\theta_J, \beta\,\}$ parameter space with SpinTaylorF2
  injections that are distributed uniformly in chirp time, $\{\tau_0,\tau_3\}$,
  with $\widehat{\mathbf{J}}$ distributed uniformly over the sphere.}
\label{fig:ff_thetabeta}
\end{figure}

To further investigate the differences between the FOB and ASB
template banks, we now calculate the difference between the fitting
factor obtained for the FOB and the ASB (we compute
$FF_{FOB}-FF_{ASB}$) and plot the result over different slices of the
parameter space.  These plots break up the relative performance of the
two banks over different portions of the parameter space.
Figs.~\ref{fig:matchdiff_tau0tau3_faceon} and
\ref{fig:matchdiff_tau0tau3} plot the difference in the fitting
factors over $(\tau_0,\tau_3)$ space.
Figs.~\ref{fig:matchdiff_qk_faceon} and \ref{fig:matchdiff_qk} show
the difference in the fitting factor in $q,\chi^\parallel$ coordinates
for face-on and arbitrary injections respectively. Here, in Figure
\ref{fig:matchdiff_qk_faceon}, as expected, we see that the FOB always
performs better. Further, in the regions where the metric has highest
density (see. Figure \ref{fig:FaceOnBankqS1L}), the FOB shows the most
improvement.  Finally, in what is possibly more illuminating,
Figure~\ref{fig:ff_thetabeta} shows the fitting factors for the FOB in
the space of $\theta_J$ and the precession cone opening angle
$\beta\,$. We quote the value of $\beta\,$ at a reference frequency of
100\,Hz.  While in principle $\beta\,$ evolves in time and thus has a
frequency dependence, it was shown in \cite{Brown:2012gs} that it is
roughly constant over the inspiral regime for the frequency range of
interest for ground based detectors. Figure~\ref{fig:ff_thetabeta}
shows a clear correlation between the spin orientation and the opening
angle.  To a good approximation, the figure shows a circle in the
$\theta_J,\beta$ plane i.e. cone around the $\beta=90^\circ$
direction. This relation was found analyticaly in \cite{Brown:2012gs}
and we refer the reader to this paper for further discussion.

To quantify the improvement that a precessing face-on bank would bring
to a Compact Binary Coalescence (CBC) search, we calculated the relative improvement in detection
volume \cite{Canton:2014ena} of the FOB and ASB banks.  In the absence
of any prior astrophysical likelihood distribution of NSBH systems,
the detector volume, $\mathcal{V}$, is proportional to the sum of the
cube of the product of the optimal SNR of the injections, $\rho_{i}$,
with the fitting factor, $m_{i}$, obtained from attempting to recover
a set of injected NSBH signals into the bank,
\begin{equation}
\mathcal{V\propto\mathrm{\sum_{i}(m_{i}\rho_{i})^{3}}} \,.\label{eq:detectorVol_propto}
\end{equation}
By taking the ratio of the detection volumes of the FOB and ASB,
$\mathcal{V_{\mathrm{FOB}}}$ vs $\mathcal{V_{\mathrm{ASB}}}$ , we get
a measure of the relative improvement the FOB could bring to
the search. Results are shown in Table \ref{tab:detector_volumes}.

\section{Conclusions }

In this paper we have presented a template bank for gravitational wave
searches for precessing NSBH systems.  The template bank assumes that
the total angular momentum vector is pointing directly towards or away
from the detectors. It covers the mass ranges $2M_\odot < M_1 < 16
M_\odot$, $1M_\odot < M_2 < 3M_\odot$ and the black-hole spin vector
can have arbitrary orientation.  The template bank ends up having
$6,908,681$ templates assuming the early Advanced LIGO noise curve.
We have shown that the sensitive volume for systems with large spin
misalignments (i.e. large precession cone angles) for this template
bank is roughly twice as large as for the aligned spin bank (see third
row of Table \ref{tab:detector_volumes}).

\begin{table*}
  \caption{Table of the improvement in the relative detection volumes calculated from each injection set. The values in the third row represent injections with the component masses distributed uniformly within their respective ranges, spins distributed uniformly in $\kappa$, and $\widehat{\mathbf{J}}$ distributed uniformly over the sphere. Results are grouped into three different regions $\{All\, ,HP\, , High\beta\, \}$. $All$ is the entire NSBH parameter space spanned by the injection set. $High\beta\,$ is defined as the region of parameter space that contains recovered injections with $\beta\,\in\,\{60^{\circ},120^{\circ}\}$. $HP$ is the ``High Precession'' region of parameter space examined by \cite{Canton:2014uja} that contains recovered injections with $||\mathbf{\chi}||>0.7$ and $45^{\circ}<\theta_{J}<135^{\circ}$.}
\label{tab:detector_volumes} 
 \begin{center}
    \begin{tabular}{  | l | c | c | c || c | c | c | } 
     \hline			
     Injected Waveform & $\theta_J$ & Mass Range $M_\odot$&
     $\frac{\mathcal{V_{\mathrm{{FOB}}}^{\mathrm{All}}}}{\mathcal{V_{\mathrm{{ASB}}}^{\mathrm{All}}}}-1$ &
     $\frac{\mathcal{V_{\mathrm{{FOB}}}^{\mathrm{HP}}}}{\mathcal{V_{\mathrm{{ASB}}}^{\mathrm{HP}}}}-1$ &
     $\frac{\mathcal{V_{\mathrm{{FOB}}}^{\mathrm{High\beta}}}}{\mathcal{V_{\mathrm{{ASB}}}^{\mathrm{High\beta}}}}-1$\\ \hline
      SpinTaylorF2 & $0^\circ$ & $\{2-16,1-3\}$ &  $3.26\%$ & $3.41\%$ & $14.2\%$\\
      SpinTaylorF2 & $0^\circ$ & $\{15,1.4\}$ &  $6.42\%$ & $4.66\%$ &$23.9\%$\\
      SpinTaylorF2 & $0-180^\circ$ & $\{2-16,1-3\}$&  $23.7\%$ & $9.88\%$ &$134\%$\\
      SpinTaylorF2 & $0-180^\circ$ & $\{15,1.4\}$ & $11.3\%$ & $3.22\%$ &$14.4\%$\\
     \hline
    \end{tabular}
  \end{center}
 \end{table*}


We use the frequency domain, inspiral-only, SpinTaylorF2 waveform for
our study. The aligned spin template bank over the same mass range has
only $130,646$ templates and this great increase in the number of
templates is validated by an independent numerical evaluation of the
determinant of the parameter space metric.  Despite this large
increase in the number of templates, we show that stochastic methods
can still be implemented.  It requires us to break up the parameter
space into smaller, approximately independent regions and we found
that the chirp times provide a suitable coordinate choice with which
to do this.  The template bank could be pruned by removing templates
near the boundaries of the chirp time boxes but this would only reduce
the number of templates by about 5-10\%.  Using a different detection
statistic as in \cite{Harry:2016ijz} should further help in decreasing
the number of templates somewhat, but it is still an open issue
whether the 97\% minimal match condition should be kept as
gravitational wave detectors improve their low frequency
sensitivity. In either case, working in chirp time coordinates should
allow us to deal with the computational problem.

A large fraction of the templates of our bank are in the anti-aligned part of
parameter space (with $\kappa < -0.5$).  If one believes that such
systems are disfavored astrophysically, it is straightforward to
construct a precessing template bank for restricted values of
$\kappa$.  Depending on how restricted we would like the black-hole
spin orientation to be, this might provide a useful compromise between
computational cost and astrophysical priors. It would also be
desirable to be able to apply traditional geometric methods and to
place a lattice of templates, but this requires us to find suitable
coordinates for the space of precessing signals.

\begin{acknowledgments}
T.D.C. is supported by an appointment to the NASA Postdoctoral Program
at the Goddard Space Flight Center, administered by Universities Space
Research Association under contract with NASA. KH acknowledges DST-MPG Max Planck Partner Group at IISER TVM for the travel support for the visit to the Albert Einstein Institute during which part of the work was carried out. This paper has LIGO document number LIGO-P1600330.
\end{acknowledgments}

\bibliographystyle{apsrev4-1}
\bibliography{stochprec}

\end{document}